\documentclass[pteplogo]{ptephy_v1}

\usepackage[hidelinks,colorlinks=true,linkcolor=blue, citecolor=blue, urlcolor=blue]{hyperref}
\usepackage{url}
\usepackage{fancybox}
\usepackage{geometry}
\usepackage{wrapfig}
\usepackage{amsmath,cases}
\usepackage{amssymb}
\usepackage{bbold}
\usepackage{mathrsfs}
\usepackage{braket}
\usepackage{longtable}
\usepackage{dcolumn}
\usepackage{multirow}
\usepackage{framed}
\usepackage{eclbkbox}
\usepackage{arydshln}
\usepackage{chngcntr}
\usepackage{booktabs}
\usepackage{here}

\newcommand{\vect}{\boldsymbol}

\newcommand{\varparallel}{\def\@varparallel{/\kern-.2em /}
\mathchoice
{\@varparallel}
{\textstyle\@varparallel} {\scriptscriptstyle\@varparallel} {\scriptscriptstyle\@varparallel}}
\DeclareSymbolFont{largesymbol}{OMX}{yhex}{m}{n}
\DeclareMathAccent{\Widehat}{\mathord}{largesymbol}{"62}

\begin{document}

\title{Implementation of chiral two-nucleon forces to nuclear many-body methods with Gaussian-wave packets}

\author{Tokuro Fukui}
\affil{Faculty of Arts and Science, Kyushu University, Fukuoka 819-0395, Japan,\\
and RIKEN Nishina Center, Wako 351-0198, Japan \email{tokuro.fukui@artsci.kyushu-u.ac.jp}}

\begin{abstract}%
 Many-body methods that use Gaussian-wave packets to describe nucleon-spatial distribution
 have been widely employed for depicting various phenomena in nuclear systems, in particular clustering. 
 So far, however, the chiral effective field theory, a state-of-the-art theory of nuclear force, 
 has not been applied to such methods. In this paper, we give the formalism to calculate the two-body matrix elements
 of the chiral two-nucleon forces using the Gaussian-wave packets. 
 We also visualize the matrix elements and investigate the contributions of the central and tensor forces.
 This work is a foothold towards an \textit{ab initio} description of various cluster phenomena
 in view of nucleons, pions, and many-nucleon forces.
\end{abstract}

\subjectindex{D00, D10, D11}

\maketitle

\section{Introduction}
\label{SecIntro}
So far, an enormous number of studies have been carried out 
to understand various phenomena in nuclear systems from the view point of nuclear force.
For this purpose, many-body methods based on the local Gaussian basis functions 
have been established and widely applied;
for example, the fermionic molecular dynamics (FMD)~\cite{FELDMEIER1990147,FELDMEIER1995493}
and antisymmetrized molecular dynamics (AMD)~\cite{HISASHI1991257,KANADAENYO2003497,10.1093/ptep/pts001}.
These methods employ the Gaussian-wave packet as 
the spatial part of the single-nucleon wave function,
and have an advantage in describing localized cluster phenomena.
However, a way to combine these methods with the chiral effective field theory 
(EFT)~\cite{Weinberg1979327,Epelbaum2006654,MACHLEIDT20111}, a state-of-the art theory of nuclear force,
has not been proposed.

The chiral EFT derives nuclear forces from low-energy quantum chromodynamics (QCD),
where nucleons and pions are the effective degree of freedom under the chiral symmetry of QCD.
The chiral EFT succeeded phenomenologically, i.e., 
it gives realistic nuclear forces that describe nucleon-nucleon phase shifts with high precision
(see Ref.~\cite{PhysRevC.96.024004} for example).
Furthermore, the chiral EFT has an advantage over previous theories in consistently handling many-nucleon forces.
Therefore, nowadays, the potentials derived from the chiral EFT are regarded as a standard input of \textit{ab initio} calculations.

The purpose of this work is to lay the foundation of implementing the chiral interaction to
the many-body method based on the Gaussian-wave packets.
We put the main focus on deriving the two-body matrix elements (MEs) of the chiral interaction by local Gaussians,
and then, investigate their behavior.
We expect that results of this work will be a milestone to understand how nuclear clusters emerge from
the fundamental degree of freedom, namely, nucleons and pions.
In this article, we adopt the two-nucleon force at next-to-next-to-next-to-leading order (N$^3$LO).
At this order, the chiral interaction becomes realistic,
i.e., the $\chi^2$/datum for the reproduction of nucleon-nucleon scattering data reaches about unity~\cite{ENTEM200293,PhysRevC.66.014002,PhysRevC.68.041001,MACHLEIDT20111}.
The higher-order contributions and many-nucleon forces will be sought in forthcoming works.

Bare nucleon-nucleon interaction, such as the chiral interaction, leads to the strong tensor and short-range correlations.
How to handle these correlations within the many-body methods with the Gaussian-wave packets has been intensively studied.
However, this point leis outside the scope of this paper, since the main target is now the formulation of the two-body MEs, as mentioned above.
Therefore, here, we just mention previous studies on the correlations with the Gaussian-wave packets.

As regards the tensor correlation, recent works with AMD~\cite{10.1093/ptep/ptx143,PhysRevC.106.044310} 
and the microscopic cluster model~\cite{PhysRevC.97.014304} showed that the application of the imaginary Gaussian center and spatially compact wave packets
to these methods is an efficient way to take into account the high-momentum components of two-nucleon pairs induced by the tensor interaction.
The formalism of this paper allows us to use complex numbers as the Gaussian center and vary the wave-packet size.

Also, there are many papers addressing how to manage the short-range correlation with the local Gaussian many-body methods.
For instance, FMD was combined with the realistic Argonne interaction~\cite{PhysRevC.51.38}
transformed into a phase-shift equivalent low-momentum interaction 
by means of the unitary correlation operator method~\cite{FELDMEIER199861,NEFF2003311,ROTH201050},
and it was applied to the investigation of nuclear clusters~\cite{NEFF2004357,PhysRevLett.98.032501}
and the simulation of the astrophysical reactions, $^{3}\mathrm{He}(\alpha,\gamma)^{7}\mathrm{Be}$ and 
$^{3}\mathrm{H}(\alpha,\gamma)^{7}\mathrm{Li}$~\cite{PhysRevLett.106.042502}.
Furthermore, the short-range correlation was investigated within the framework of AMD~\cite{10.1093/ptep/pty020}
and the quasi-cluster model~\cite{10.1093/ptep/ptz046}
by directly using nucleon-nucleon potentials with a repulsive core.
In future, the finding by these studies will be helpful to clarify the role of the short-range correlation induced by the chiral interaction
within the local Gaussian many-body methods.

We put a comment on the first attempt to combine the cluster model and the chiral interaction.
In our recent study~\cite{Fukui_2022}, the effective interaction relevant to the Brink model~\cite{brink1966proc}
was obtained from the chiral interaction, but the noncentral forces were all missing due to presumed $\alpha$ clusters
and a phenomenological prescription was introduced.
Such shortcomings will be overcome by an \textit{ab initio} calculation
based on the formalism of this work.

This article is constructed as follows.
In Sec.~\ref{SecMEs}, we give the formalism of the two-body MEs in momentum space.
In Sec.~\ref{SecVis}, as an example, typical MEs are visualized.
Then, Sec.~\ref{SecConc} is devoted to a summary and future perspectives.
Appendix~\ref{SecMBWFProj} briefly shows how the MEs formulated in this work enter the practical many-body calculations.
The detail of the chiral potentials and information relevant to the formalism are relegated to Appendices~\ref{SecChiralpot} and~\ref{SecMPE}.


\section{Formalism}
\label{SecMEs}
\subsection{Wave functions}
\label{SecWF}
We compute the two-body MEs in momentum space for the chiral interaction as the two-nucleon force,
while other terms of the Hamiltonian can be calculated analytically (see Ref.~\cite{brink1966proc} for example).
Since the main focus of this paper is the derivation of the two-body MEs of the chiral interaction by local Gaussians,
we express the single-particle wave function in momentum space:
\begin{align}
 \Ket{\varphi_i^{(\nu)}}
 &=
 \Ket{\phi_i^{(\nu)}\chi_i},
 \label{spwfcomplex1}\\
 \Braket{\vect{p}_n|\phi_i^{(\nu)}}
 &=
 \phi_i^{(\nu)}(\vect{p}_n)
 =
 \frac{1}{\left(2\pi\nu\right)^{\frac{3}{4}}}
 \exp\!\left[-\frac{1}{4\nu}(\vect{p}_n+2i\nu\vect{Z}_i)^2 \right]\exp\!\left[-\nu Z_i^2\right],
 \label{spwfcomplex2}\\
 \Ket{\chi_i}
 &=
 \left[\alpha_i\Ket{\uparrow_{\sigma}}+\beta_i\Ket{\downarrow_{\sigma}}\right]
 \Ket{m_{\tau_i}},
 \label{spinisospinwf}
\end{align}
where $\phi_i^{(\nu)}$ is the Fourier transform of its coordinate-space representation:
\begin{align}
 \phi_i^{(\nu)}(\vect{p}_n)
 =
 \frac{1}{(2\pi)^{\frac{3}{2}}}
 \left(\frac{2\nu}{\pi}\right)^{\frac{3}{4}}
 \int d\vect{r} \exp\!\left(-i\vect{p}_n\cdot\vect{r}\right)
 \exp\!\left[-\nu\left(\vect{r}-\vect{Z}_i\right)^2\right],
 \label{FourierSPWF}
\end{align}
with $\vect{p}_n$ the momentum of the $n$th nucleon.
In this paper we use natural units such that $\hbar=c=1$.
The Gaussian-wave packet is characterized by the range parameter $\nu$ 
and the Gaussian center $\vect{Z}_i$, which is in general complex.
The spin-isospin state $\Ket{\chi_i}$ is expressed as a superposition
of the spin-up and spin-down states, $\Ket{\uparrow_{\sigma}}$ and $\Ket{\downarrow_{\sigma}}$, respectively.
The weights $\alpha_i$ and $\beta_i$ are determined by many-body calculations through variational processes.
The isospin state can be $\Ket{m_{\tau}}=\Ket{\uparrow_{\tau}}$ (neutron) or 
$\Ket{m_{\tau}}=\Ket{\downarrow_{\tau}}$ (proton).

As argued in Refs.~\cite{10.1093/ptep/ptx143,PhysRevC.97.014304},
the imaginary part of $\vect{Z}_i$ corresponds to the expectation value of the nucleon momentum.
Indeed, if $\vect{Z}_i$ is pure imaginary in Eq.~\eqref{spwfcomplex2}, 
$\phi_i^{(\nu)}$ localizes at around $\vect{p}_n\sim 2\nu \mathrm{Im}(\vect{Z}_i)$.
Therefore, complex $\vect{Z}_i$ enables us to efficiently take into account the high-momentum components
caused by the tensor correlation.
Also, inclusion of the spatially compact wave packets by varying $\nu$ can improve calculations~\cite{PhysRevC.106.044310}.
Our formalism allows to vary $\nu$ and we explain it in Appendix~\ref{SecMBWFProj}, 
where many-body states are defined using the single-particle wave function of Eq.~\eqref{spwfcomplex1}.

From Eq.~\eqref{spwfcomplex1}, the two-body state in momentum space reads
\begin{align}
 \Braket{ \vect{p}_1, \vect{p}_2 | \phi_i^{(\nu)}\phi_j^{(\nu)}}
 =
 \frac{1}{\left(2\pi\nu\right)^{\frac{3}{2}}}
 \exp\!\left[-\frac{1}{4\nu}\left\{(\vect{p}_1+2i\nu\vect{Z}_i)^2+(\vect{p}_2+2i\nu\vect{Z}_j)^2\right\} -\nu \left(Z_i^2+Z_j^2\right)\right],
 \label{2BWFcomplex}
\end{align}
which can be rewritten in terms of the center-of-mass (CM) momentum $\vect{P}$
and the relative momentum $\vect{p}$ as
\begin{align}
 \Braket{ \vect{p}_1, \vect{p}_2 | \phi_i^{(\nu)}\phi_j^{(\nu)}}
 &=
 \frac{1}{\left(2\pi\nu\right)^{\frac{3}{2}}}
 \exp\!\left[-\frac{P^2}{8\nu}-i\vect{Z}_{ij}\cdot\vect{P}\right]
 \exp\!\left[-\frac{p^2}{2\nu}-i\vect{z}_{ij}\cdot\vect{p}\right],
 \label{2BWFcomplexRelCM}
\end{align}
with
\begin{align}
  \begin{pmatrix}
  \vect{Z}_{ij} \\
  \vect{z}_{ij}
 \end{pmatrix}
 =
 \mathcal{U}
 \begin{pmatrix}
  \vect{Z}_i \\
  \vect{Z}_j
 \end{pmatrix},
 \qquad
 \begin{pmatrix}
  \vect{P} \\
  \vect{p}
 \end{pmatrix}
 =
 \left(\mathcal{U}^{-1}\right)^T
 \begin{pmatrix}
  \vect{p}_1 \\
  \vect{p}_2
 \end{pmatrix},
 \qquad
 \mathcal{U}
 =
 \begin{pmatrix}
  \frac{1}{2} & \frac{1}{2} \\
  1           & -1
 \end{pmatrix}.
 \label{Umat1}
\end{align}
Here, neutrons and protons are assumed to have common mass.

\subsection{General form of two-body matrix elements}
\label{SecCompGen2BME}
In this section we derive the general expression of the two-body MEs in momentum space
with a nonlocal potential dependent on the relative momenta, $\vect{p}$ and $\vect{p}'$, of the initial and final channels, respectively.
The nonlocality originates from the regularization scheme of the chiral EFT (see Appendix~\ref{SecChiralpot}).

For simplicity, the nucleon spin is assumed to be either $\Ket{\uparrow_{\sigma}}$ or $\Ket{\downarrow_{\sigma}}$.
Note that this does not undermine the loss of generality of the formalism.
Thus, we derive the MEs based on the single-particle state $\Ket{\phi_i^{(\nu)}\xi_i}$,
where $\phi_i^{(\nu)}$ is defined by Eq.~\eqref{spwfcomplex2} and $\Ket{\xi_i}$ is now given by
\begin{align}
 \Ket{\xi_i}
 &=
 \Ket{m_{\sigma_i}}
 \Ket{m_{\tau_i}}.
 \label{spinisospinwfalign}
\end{align}
Hence, the spin-up and spin-down states are respectively expressed by $\Ket{m_{\sigma_i}}=\Ket{\uparrow_{\sigma}}$
and $\Ket{m_{\sigma_i}}=\Ket{\downarrow_{\sigma}}$.
Now the two-body states are represented by 
\begin{align}
 \Ket{i j,\nu}=\Ket{\phi_i^{(\nu)}\phi_j^{(\nu)}\xi_i\xi_j},
 \label{2bstatesalign}
\end{align}
with which the MEs are formulated.

Let us move to the general form of the two-body MEs.
A general two-body operator $\hat V_{2N}$ can be written as
\begin{align}
 \hat V_{2N}
 &=
 \int\!\!\!\!\int\!\!\!\!\int\!\!\!\!\int
 d\vect{p} d\vect{p}' d\vect{P} d\vect{P}'
 \Ket{\vect{p}',\vect{P}'}
 \left< \vect{p}', \vect{P}' \left| \left. \hat V_{2N}\right. \right| \vect{p}, \vect{P} \right>
 \Bra{\vect{p},\vect{P}}.
 \label{OpeV2N1}
\end{align}
The Galilean invariance requires that 
the potential $v_{2N}$ in association with the operator $\hat V_{2N}$ does not depend on
$\vect{P}$ and $\vect{P}'$.
Hence, such a potential is given by
\begin{align}
 \left< \vect{p}', \vect{P}' \left| \left. \hat V_{2N}\right. \right|
 \vect{p}, \vect{P} \right>
 =
 v_{2N}\!\left(\vect{p}',\vect{p}\right)
 \delta(\vect{P}-\vect{P}'),
 \label{genepot1}
\end{align}
which results in the two-body MEs:
\begin{align}
 &\left< kl,\nu' \left| \left. \hat V_{2N}\right. \right| ij,\nu \right>
 \nonumber\\
 &\quad=
 \int\!\!\!\!\int\!\!\!\!\int
 d\vect{p} d\vect{p}' d\vect{P}
 \left< kl,\nu' \left| \vect{p}', \vect{P} \right>\right.
 v_{2N}\!\left(\vect{p}',\vect{p}\right)
 \left< \vect{p}, \vect{P} \left| ij,\nu \right>\right.
 \nonumber\\
 &\quad=
 A_{ijkl}^{(\nu\nu')}
 \int\!\!\!\!\int
 d\vect{p} d\vect{p}'
 \left< \xi_k\xi_l \left| \left. v_{2N}\!\left(\vect{p}',\vect{p}\right) \right. \right| \xi_i\xi_j \right>
 g_{\nu\nu'}(p,p')
 \exp\!\left[-i\vect{z}_{ij}\cdot\vect{p}+i\vect{z}_{kl}^*\cdot\vect{p}'\right],
 \label{gene2bMEmom2Complex}
\end{align}
with
\begin{align}
 g_{\nu\nu'}(p,p')
 &=
 \exp\!\left[-\frac{p^2}{2\nu}-\frac{p'^2}{2\nu'}\right],
 \label{gaussgnunu'}\\
 A_{ijkl}^{(\nu\nu')}
 &=
 \left[\frac{2}{\pi(\nu+\nu')}\right]^{\!\frac{3}{2}}
 \exp\!\left[-\frac{2\nu\nu'}{\nu+\nu'}\left(\vect{Z}_{ij}-\vect{Z}_{kl}^*\right)^2\right].
 \label{Aijklnunu'}
\end{align}
Using Eq.~\eqref{gene2bMEmom2Complex},
we introduce the notation of the antisymmetrized two-body MEs,
\begin{align}
 V_{ijkl}^{(\nu\nu')}=
 \left< kl,\nu' \left| \left. \hat V_{2N}\right. \right| ij,\nu \right>
-\left< kl,\nu' \left| \left. \hat V_{2N}\right. \right| ji,\nu \right>.
 \label{antisym2BMEs}
\end{align}

For later convenience, we perform the Rayleigh expansion for the plane waves in Eq.~\eqref{gene2bMEmom2Complex}:
\begin{align}
 \exp\!\left[-i\vect{z}_{ij}\cdot\vect{p}\right]
 &=
 4\pi \sum_{L} (-)^L i^{L}
 j_{L}\!\left(z_{ij}p\right)
 \sum_{M}(-)^M
 Y_{LM}\!\left(\hat{\vect{p}}\right) Y_{L,-M}\!\left(z_{ij,x},z_{ij,y},z_{ij,z}\right),
 \label{Rayleighcomplex1}\\
 \exp\!\left[i\vect{z}_{kl}^*\cdot\vect{p}'\right]
 &=
 4\pi \sum_{L'} i^{L'}
 j_{L'}\!\left(z_{kl}^*p'\right)
 \sum_{M'}
 Y_{L'M'}^*\!\left(\hat{\vect{p}}'\right) Y_{L'M'}\!\left(z_{kl,x}^*,z_{kl,y}^*,z_{kl,z}^*\right),
 \label{Rayleighcomplex2}
\end{align}
where $\vect{z}_{ij}=(z_{ij,x},z_{ij,y},z_{ij,z})$.
Note that $z_{ij,x}$, $z_{ij,y}$, and $z_{ij,z}$ are complex numbers.
Hence, the arguments of the spherical Bessel function $j_L$ are also complex numbers as 
${z_{ij}=\sqrt{\vect{z}_{ij}\cdot\vect{z}_{ij}}}$ and $z_{kl}^*=\sqrt{\vect{z}_{kl}^*\cdot\vect{z}_{kl}^*}$.
The spherical harmonics $Y_{LM}$ with the complex numbers as its arguments is given by the analytic continuation in the Cartesian representation:
\begin{align}
 Y_{lm}(z_x,z_y,z_z)
 &=
 (-)^m \sum_{\lambda=0}^{[(l-m)/2]}
 \left[\frac{(2l+1)(l-m)!}{4\pi(l+m)!}\right]^{\frac{1}{2}} \frac{(l+m)!}{2^m}
 \frac{(-)^\lambda}{4^\lambda\lambda!(m+\lambda)!(l-m-2\lambda)!}
 \nonumber\\
 &\times
 \left(\frac{z_z}{z}\right)^{l-m-2\lambda}
 \left(\frac{z_x+iz_y}{z}\right)^{m+\lambda}
 \left(\frac{z_x-iz_y}{z}\right)^\lambda
 \qquad (m\geq 0).
 \label{SHcomplex}
\end{align}
The upper limit of the summation is expressed by the floor function
and ${z=\sqrt{\vect{z}\cdot\vect{z}}=\sqrt{z_x^2+z_y^2+z_z^2}}$ is a complex number.

\subsection{Two-body matrix elements of chiral interaction}
\label{Sec2BMEchiral}
\subsubsection{Overview}
\label{Sec2BMEchiraloverview}
Now we formulate the two-body MEs of the chiral interaction at N$^3$LO.
One finds that the chiral-N$^3$LO potential $v_{2N}$ 
consists of the central, spin-orbit (SO), and tensor terms.
Therefore, the MEs can also be decomposed into these terms as 
$V_{ijkl}^{(\nu\nu')}=V_{ijkl}^{(\nu\nu':C)}+V_{ijkl}^{(\nu\nu':LS)}+V_{ijkl}^{(\nu\nu':T)}+V_{ijkl}^{(\nu\nu':\sigma L)}$,
where the central, SO, tensor, and $\sigma L$ contributions are denoted by
$V_{ijkl}^{(\nu\nu':C)}$, $V_{ijkl}^{(\nu\nu':LS)}$, $V_{ijkl}^{(\nu\nu':T)}$, $V_{ijkl}^{(\nu\nu':\sigma L)}$, respectively.
Although the $\sigma L$ term, the potential of which depends 
on $V_{\sigma L}^{(4)}$ and $D_{15}$ in Eqs.~\eqref{potN3LO2pi} and~\eqref{potN3LOct}, respectively,
is one of the tensor contributions,
for simplicity, we treat it separately from standard tensor terms.

In the following sections, first we derive the typical MEs of the central, SO, and tensor contributions to explain basic procedures of the calculations.
The MEs of the contact terms at next-to-leading order (NLO) are chosen as examples
since they consist of the central, SO, and tensor components, and
their MEs are relatively simple due to the absence of the pion propagators.
Then, unified expressions of the MEs are given in Sec.~\ref{Sec2BMEchiralsummary}.

\subsubsection{Central contributions}
\label{Sec2BMEchiralcentral}
As an example of the central contributions, we take $C_1$ term,
one of the contact terms at NLO. 
As shown by Eq.~\eqref{potNLOct}, with the nonlocal regularization, its potential is given by
\begin{align}
 v_{2N}\!\left(\vect{p}',\vect{p}\right)
 &=
 \frac{1}{(2\pi)^3}C_1 q^2 u_n\!\left(p,p',\Lambda\right),
 \label{potC1}
\end{align}
where $C_1$ is one of the low-energy constants (LECs). 
The transferred momentum $\vect{q}$, as well as the average momentum $\vect{Q}$ are defined by
\begin{align}
  \begin{pmatrix}
  \vect{Q} \\
  \vect{q}
 \end{pmatrix}
 =
 \mathcal{U}
 \begin{pmatrix}
  \vect{p}' \\
  \vect{p}
 \end{pmatrix},
 \label{UtransppqQmain}
\end{align}
with the matrix $\mathcal{U}$ given by Eq.~\eqref{Umat1}.
The regulator $u_n$ characterized by the power $n$ and the cutoff momentum $\Lambda$
is nonlocal:
\begin{align}
 u_n\!\left(p,p',\Lambda\right)
 &=
 \exp\!\left[-\!\left(\frac{p}{\Lambda}\right)^{\!2n}-\!\left(\frac{p'}{\Lambda}\right)^{\!2n}\right].
 \label{nonlocalregmain}
\end{align}
See Appendix~\ref{SecChiralpot} for more detail of the potential form.
To compute the MEs, we perform the multipole expansion (MPE) of $q^2$ in Eq.~\eqref{potC1}:
\begin{align}
 &q^2
 =
 4\pi\sum_{K=0,1}
 \frac{(-)^K}{\hat{K}}\mathcal{F}^{(C_1)}_K\!\left(p,p'\right)
 \left[Y_K\!\left(\hat{\vect{p}}\right)\otimes Y_K\!\left(\hat{\vect{p}}'\right)\right]_{00},
 \label{q2YY2}
\end{align}
and the MPE function is given as
\begin{align}
 \mathcal{F}^{(C_1)}_K\!\left(p,p'\right)
 =
 \left(p^2+p'^2\right)\delta_{K 0}
 -2pp'\delta_{K 1}.
 \label{q2MPEfun}
\end{align}
The bipolar spherical harmonics is defined by
\begin{align}
 \left[Y_L\!\left(\hat{\vect{p}}\right)\otimes Y_{L'}\!\left(\hat{\vect{p}}'\right)\right]_{\lambda\mu}
 =
 \sum_{M=-L}^{L} \sum_{M'=-L'}^{L'}
 \left( L M L' M' | \lambda \mu \right)
 Y_{LM}\!\left(\hat{\vect{p}}\right) Y_{L'M'}\!\left(\hat{\vect{p}}'\right).
 \label{bipolarYY}
\end{align}
We adopt the abbreviation $\hat{L}=\sqrt{2L+1}$
and the Clebsch-Gordan coefficient is denoted by
$(\cdots\cdot\cdot\cdot|\cdot\cdot)$.
In Appendix~\ref{SecMPE}, the detail of the MPE can be found.

Using these expressions and plug Eq.~\eqref{potC1} into Eq.~\eqref{gene2bMEmom2Complex},
as well as employing Eqs.~\eqref{Rayleighcomplex1} and~\eqref{Rayleighcomplex2},
one finds the antisymmetrized MEs as
\begin{align}
 V_{ijkl}^{(\nu\nu':C)}
 &=
 \frac{2}{\pi} C_1
 A_{ijkl}^{(\nu\nu')}
 \sum_{L=0,1}
 \left[D_{ijkl}^{(\sigma)} D_{ijkl}^{(\tau)}-(-)^LD_{jikl}^{(\sigma)} D_{jikl}^{(\tau)}\right]
 P_L(z_{ijkl})
 \nonumber\\
 &\times
 \int\!\!\!\!\int
 dp dp'
 p^{2}p'^{2}
 g_{\nu\nu'}(p,p') u_n\!\left(p,p',\Lambda\right)
 j_{L}\!\left(z_{ij}p\right)  j_{L}\!\left(z_{kl}^*p'\right)
 \mathcal{F}^{(C_1)}_L\!\left(p,p'\right).
 \label{antisymC1ME}
\end{align}
Here, the Legendre polynomial $P_{L}$ has the argument given by
\begin{align}
 z_{ijkl}
 =
 \left\{
 \begin{aligned}
  &1 && \qquad (\text{$i=j$ and/or $k=l$}),\\
  &\frac{\vect{z}_{ij}\cdot\vect{z}_{kl}^*}{z_{ij}z_{kl}^*} && \qquad (\text{$i\ne j$ and $k\ne l$}),
 \end{aligned}
 \right.
 \label{zijklargument}
\end{align}
and $D_{ijkl}^{(\sigma)}$ is defined by
\begin{align}
 D_{ijkl}^{(\sigma)}
 &=
 \delta_{m_{\sigma_i} m_{\sigma_k}}
 \delta_{m_{\sigma_j} m_{\sigma_l}}.
 \label{Dijkl}
\end{align}
The superscripts $(\sigma)$ and $(\tau)$ stand for the spin and isospin MEs, respectively.

We emphasize that any central term of the chiral potential can be represented
in terms of $\left[Y_K\!\left(\hat{\vect{p}}\right)\otimes Y_K\!\left(\hat{\vect{p}}'\right)\right]_{00}$
as Eq.~\eqref{q2YY2}.
This is because the spatial and spin parts of the central operators are decoupled,
i.e., they both form a scalar operator individually.
For the pion-exchange terms, the MPE of the $q$- and $Q$-dependent parts of the potentials
involving the pion propagators is necessary, but after some manipulation, 
$\left[Y_K\!\left(\hat{\vect{p}}\right)\otimes Y_K\!\left(\hat{\vect{p}}'\right)\right]_{00}$
must appear.
As a result, the structure of the central MEs is always expressed by Eq.~\eqref{antisymC1ME}.
Indeed, for other central terms, we just have to replace the prefactor 
$2 C_1/\pi$ and $\mathcal{F}^{(C_1)}_L$ with appropriate forms, 
as explained in Sec.~\ref{Sec2BMEchiralsummary} and Appendix~\ref{SecMPE}.
Note that, in principle, the summation over $L$ runs up to infinity for general cases,
and the spin-isospin MEs need to be modified depending on the operator forms (see Table~\ref{tableChiralSpinIso}).

\subsubsection{Spin-orbit contributions}
\label{Sec2BMEchiralSO}
The $C_5$ term, one of the contact terms at NLO is a good example of the SO contributions.
Its potential reads
\begin{align}
 v_{2N}\!\left(\vect{p}',\vect{p}\right)
 &=
 \frac{1}{(2\pi)^3}C_5 u_n\!\left(p,p',\Lambda\right)
 \left[-i\vect{S}\cdot\left(\vect{q}\times\vect{Q}\right)\right],
 \label{potC5}
\end{align}
where $C_5$ is the LEC. The total spin $\vect{S}$ is given by
\begin{align}
 \vect{S}&=\frac{1}{2}\left(\vect{\sigma}_1+\vect{\sigma}_2\right).
 \label{totalS}
\end{align}
The spin operator $\vect{\sigma}_i$ is represented by the Pauli matrices.
The operator $-i\vect{S}\cdot\left(\vect{q}\times\vect{Q}\right)$ can be represented 
in terms of $\vect{p}$ and $\vect{p}'$ as
\begin{align}
 -i\vect{S}\cdot\left(\vect{q}\times\vect{Q}\right)
 &=
 4\pi\sqrt{\frac{3}{2}}
 \sum_{\lambda_q=0,1}
 \sum_{\lambda_Q=0,1}
 (-)^{\lambda_q}\widehat{1\!-\!\lambda_q}\widehat{1\!-\!\lambda_Q}
 \left[\binom{3}{2\lambda_q}\binom{3}{2\lambda_Q}\right]^{\!\frac{1}{2}}
 p^{\lambda_q+\lambda_Q} p'^{\,2-\lambda_q-\lambda_Q}
 \nonumber\\
 &\times
 \sum_{K_1K_2}
 \left( \lambda_q 0 \lambda_Q 0 | K_1 0 \right)
 \left( 1\!-\!\lambda_q, 0, 1\!-\!\lambda_Q, 0 | K_2 0 \right)
 \begin{Bmatrix}
  \lambda_q & 1\!-\!\lambda_q & 1 \\
  \lambda_Q & 1\!-\!\lambda_Q & 1 \\
  K_1       & K_2         & 1
 \end{Bmatrix}
 \nonumber\\
 &\times
 \left[
 \left[Y_{K_1}\!\left(\hat{\vect{p}}\right)\otimes
 Y_{K_2}\!\left(\hat{\vect{p}}'\right)\right]_{1}
 \otimes S_1
 \right]_{00},
 \label{C5ope}
\end{align}
where the binomial coefficient is given by
\begin{align}
 \binom{2n+1}{2m}
 =
 \frac{\left(2n+1\right)!}{\left(2n+1-2m\right)!\left(2m\right)!},
 \label{binomialcoeff}
\end{align}
and the $9$-$j$ symbol is represented by the $3\times3$ matrix in the braces.

From Eqs.~\eqref{gene2bMEmom2Complex},~\eqref{potC5}, and~\eqref{C5ope},
the antisymmetrized MEs can be calculated as
\begin{align}
 V_{ijkl}^{(\nu\nu':LS)}
 &=
 4\sqrt{3} C_5 A_{ijkl}^{(\nu\nu')}
 \sum_{\lambda_q\lambda_Q}
 \sum_{L=0}^\infty \sum_{L'=0}^\infty
 (-)^{\lambda_Q} i^{L+L'}
 \widehat{1\!-\!\lambda_q}\widehat{1\!-\!\lambda_Q}
 \left[\binom{3}{2\lambda_q}\binom{3}{2\lambda_Q}\right]^{\!\frac{1}{2}}
 \nonumber\\
 &\quad\times
 S_{ijkl}\left[D_{ijkl}^{(\tau)}-(-)^LD_{jikl}^{(\tau)}\right]
 \nonumber\\
 &\quad\times
 \left[Y_{L}\!\left(z_{ij,x},z_{ij,y},z_{ij,z}\right) \otimes Y_{L'}\!\left(z_{kl,x}^*,z_{kl,y}^*,z_{kl,z}^*\right)\right]_{1,m_{\sigma_i}+m_{\sigma_j}-m_{\sigma_k}-m_{\sigma_l}}
 \nonumber\\
 &\quad\times
 \left( \lambda_q 0 \lambda_Q 0 | L 0 \right)
 \left( 1\!-\!\lambda_q, 0, 1\!-\!\lambda_Q, 0 | L' 0 \right)
 \begin{Bmatrix}
  \lambda_q & 1\!-\!\lambda_q & 1 \\
  \lambda_Q & 1\!-\!\lambda_Q & 1 \\
  L         & L'          & 1
 \end{Bmatrix}
 \nonumber\\
 &\quad\times
 \int\!\!\!\!\int
 dp dp'
 p^{\lambda_q+\lambda_Q+2}p'^{4-\lambda_q-\lambda_Q}
 g_{\nu\nu'}(p,p') u_n\!\left(p,p',\Lambda\right)
 j_{L}\!\left(z_{ij}p\right) j_{L'}\!\left(z_{kl}^*p'\right),
 \label{antisymC5ME}
\end{align}
with
\begin{align}
 S_{ijkl}
 &=
 \left(\left. \frac{1}{2} m_{\sigma_i} \frac{1}{2} m_{\sigma_j} \right| 1, m_{\sigma_i}\!+m_{\sigma_j} \right)
 \left(\left. \frac{1}{2} m_{\sigma_k} \frac{1}{2} m_{\sigma_l} \right| 1, m_{\sigma_k}\!+m_{\sigma_l} \right)
 \nonumber\\
 &\times
 \left( 1, m_{\sigma_i}\!+m_{\sigma_j}\!-m_{\sigma_k}\!-m_{\sigma_l}, 1,
 m_{\sigma_k}\!+m_{\sigma_l} | 1, m_{\sigma_i}\!+m_{\sigma_j} \right).
 \label{Sijkl}
\end{align}
The expression, Eq.~\eqref{C5ope}, can be applicable to other spin-orbit terms,
i.e., all spin-orbit potentials can be written in terms of
$\left[
 \left[Y_{K_1}\!\left(\hat{\vect{p}}\right)\otimes
 Y_{K_2}\!\left(\hat{\vect{p}}'\right)\right]_{1}
 \otimes S_1
 \right]_{00}$
even if pions are exchanged.
Consequently, for the MEs of the spin-orbit contributions, the structure of Eq.~\eqref{antisymC5ME} is rather general
as shown in Sec.~\ref{Sec2BMEchiralsummary}. 
In general cases, several prefactors and angular-momentum-coupling coefficients in Eq.~\eqref{antisymC5ME}
are packed into a single function,
which is the MPE function of the SO term described in Sec.~\ref{SecMPESO}.

Note that Eq.~\eqref{antisymC5ME} can be further simplified
since only $L=L'=1$ is allowed and other configurations are forbidden
by the angular-momentum-coupling coefficients involved.
However, we do not show explicitly such simplified MEs
because this section is intended to demonstrate the derivation of the SO MEs
and the general form, Eq.~\eqref{antisymC5ME}, is useful rather than showing such specific MEs.

\subsubsection{Tensor contributions}
\label{Sec2BMEchiraltensor}
The typical tensor contribution appears as the $C_6$-contact term at NLO.
As shown in Appendix~\ref{SecChiralpotNLO}, its potential is written as
\begin{align}
 v_{2N}\!\left(\vect{p}',\vect{p}\right)
 &=
 \frac{1}{(2\pi)^3}C_6u_n\!\left(p,p',\Lambda\right)
 \left(\vect{\sigma}_1\cdot\vect{q}\right)\left(\vect{\sigma}_2\cdot\vect{q}\right),
 \label{potC6}
\end{align} 
where $C_6$ is the LEC.
We express the operator in Eq.~\eqref{potC6} by the
irreducible-tensor representation:
\begin{align}
 \left(\vect{\sigma}_1\cdot\vect{q}\right)\left(\vect{\sigma}_2\cdot\vect{q}\right)
 &=
 4\pi \sum_{\lambda_0=0,2}\sum_{\lambda_1=0}^{\lambda_0}
 \frac{(-)^{\lambda_1}}{\hat{\lambda}_1}
 \binom{2\lambda_0+1}{2\lambda_1}^{\!\frac{1}{2}}
 \left( 1 0 1 0 | \lambda_0 0 \right)
 q^{2-\lambda_0} p^{\lambda_1}p'^{\lambda_0-\lambda_1}
 \nonumber\\
 &\times
 \left[
 \left[Y_{\lambda_1}\!\left(\hat{\vect{p}}\right)\otimes
 Y_{\lambda_0-\lambda_1}\!\left(\hat{\vect{p}}'\right)\right]_{\lambda_0}
 \otimes \left[\sigma_1(1)\otimes\sigma_1(2)\right]_{\lambda_0}
 \right]_{00}
 \nonumber\\
 &=
 4\pi\sum_{\lambda_0\lambda_1}\sum_{K=0,1}
 \binom{2\lambda_0+1}{2\lambda_1}^{\!\frac{1}{2}}
 \widehat{\lambda_0\!-\!\lambda_1} \left( 1 0 1 0 | \lambda_0 0 \right)
 \mathcal{F}_{\lambda_0 K}^{(C_6)}\!\left(p,p'\right)
 p^{\lambda_1}p'^{\lambda_0-\lambda_1}
 \nonumber\\
 &\times
 \sum_{K_1K_2}(-)^{K_2}
 \left( \lambda_1 0 K 0 | K_1 0\right)
 \left( \lambda_0\!-\!\lambda_1, 0 K 0 | K_2 0\right)
 \begin{Bmatrix}
  \lambda_0\!-\!\lambda_1 & \lambda_1 & \lambda_0 \\
  K_1                 & K_2       & K
 \end{Bmatrix}
 \nonumber\\
 &\times
 \left[
 \left[Y_{K_1}\!\left(\hat{\vect{p}}\right)\otimes Y_{K_2}\!\left(\hat{\vect{p}}'\right)\right]_{\lambda_0}
 \otimes
 \left[\sigma_{1}(1)\otimes \sigma_{1}(2)\right]_{\lambda_0}
 \right]_{00},
 \label{sigqsigqC6}
\end{align}
with
\begin{align}
 \mathcal{F}_{\lambda_0 K}^{(C_6)}\!\left(p,p'\right)
 =
 \left[
 \left(p^2+p'^2\right)\delta_{K 0}-2pp'\delta_{K 1}
 \right]\delta_{\lambda_0 0}
 +
 \delta_{\lambda_0 2}\delta_{K 0}.
 \label{MPEfunC6}
\end{align}
The $2\times3$ matrix in the braces is the $6$-$j$ symbol.

One finds that Eq.~\eqref{sigqsigqC6} with $\lambda_0=0$ corresponds to the central component, 
$q^2 \left(\vect{\sigma}_1\cdot\vect{\sigma}_2\right)/3$,
while that with $\lambda_0=2$ is the purely tensor component, $q^2\mathcal{S}_{12}/3$,
with the tensor operator
\begin{align}
 \mathcal{S}_{12}\!\left(\hat{\vect{q}}\right)
 &=
 3\left(\vect{\sigma}_1\cdot\hat{\vect{q}}\right)
 \left(\vect{\sigma}_2\cdot\hat{\vect{q}}\right)
 -\vect{\sigma}_1\cdot\vect{\sigma}_2
 \nonumber\\
 &=
 2\left[3\left(\vect{S}\cdot\hat{\vect{q}}\right)^2
 -\vect{S}^2\right].
 \label{tensope1}
\end{align}
Note that $\vect{S}$ is defined by Eq.~\eqref{totalS}.
In this paper, the terms dependent on $\left(\vect{\sigma}_1\cdot\vect{q}\right)\left(\vect{\sigma}_2\cdot\vect{q}\right)$
are referred to as the tensor contributions, although they involve the central contributions.

From the above expressions, the antisymmetrized MEs are computed as
\begin{align}
 V_{ijkl}^{(\nu\nu':T)}
 &=
 24 C_6 A_{ijkl}^{(\nu\nu')}
 \sum_{\lambda_0\lambda_1}
 \binom{2\lambda_0+1}{2\lambda_1}^{\!\frac{1}{2}}
 \frac{\widehat{\lambda_0\!-\!\lambda_1}}{\hat{\lambda}_0}\left( 1 0 1 0 | \lambda_0 0 \right)
  \nonumber\\
 &\times
 \sum_{LL'K} i^{L+L'}
 \left[T_{ijkl}^{(\lambda_0)} D_{ijkl}^{(\tau)}-(-)^LT_{jikl}^{(\lambda_0)} D_{jikl}^{(\tau)}\right]
 \nonumber\\
 &\times
 \left[Y_{L}\!\left(z_{ij,x},z_{ij,y},z_{ij,z}\right) \otimes Y_{L'}\!\left(z_{kl,x}^*,z_{kl,y}^*,z_{kl,z}^*\right)\right]_{\lambda_0,m_{\sigma_i}+m_{\sigma_j}-m_{\sigma_k}-m_{\sigma_l}}
 \nonumber\\
 &\times
 \left( \lambda_1 0 K 0 | L 0\right)
 \left( \lambda_0\!-\!\lambda_1, 0 K 0 | L' 0\right)
 \begin{Bmatrix}
  \lambda_0\!-\!\lambda_1 & \lambda_1 & \lambda_0 \\
  L                 & L'       & K
 \end{Bmatrix}
 \nonumber\\
 &\times
 \int\!\!\!\!\int
 dp dp'
 p^{\lambda_1+2}p'^{\lambda_0-\lambda_1+2}
 g_{\nu\nu'}(p,p') u_n\!\left(p,p',\Lambda\right)
 j_{L}\!\left(z_{ij}p\right)  j_{L'}\!\left(z_{kl}^*p'\right)
 \mathcal{F}_{\lambda_0 K}^{(C_6)}\!\left(p,p'\right),
 \label{antisymC6ME}
\end{align}
with
\begin{align}
 T_{ijkl}^{(\lambda_0)}
 &=
 (-)^{m_{\sigma_i}+m_{\sigma_j}-m_{\sigma_k}-m_{\sigma_l}}
 \left( 1, m_{\sigma_i}\!-m_{\sigma_k}, 1, m_{\sigma_j}\!-m_{\sigma_l} | 
 \lambda_0, m_{\sigma_i}\!+m_{\sigma_j}\!-m_{\sigma_k}\!-m_{\sigma_l} \right)
 \nonumber\\
 &\times
 \left(\left. \frac{1}{2} m_{\sigma_i} 1, m_{\sigma_k}\!-m_{\sigma_i} \right| \frac{1}{2} m_{\sigma_k} \right)
 \left(\left. \frac{1}{2} m_{\sigma_j} 1, m_{\sigma_l}\!-m_{\sigma_j} \right| \frac{1}{2} m_{\sigma_l} \right).
 \label{Tijkl}
\end{align}
There is no counterpart of $T_{ijkl}^{(\lambda_0)}$ for
the isospin indices, and therefore, the superscript $(\sigma)$ is not
necessary in Eq.~\eqref{Tijkl}.

The point of the calculations of the tensor MEs is that the irreducible-tensor representation
by Eq.~\eqref{sigqsigqC6} is valid also for other tensor terms
no matter whether they have the pion propagators or the operator is given by $\vect{Q}$ instead of $\vect{q}$.
Every tensor potential can be expressed in terms of 
$
 \big[\!
 \left[Y_{K_1}\!\left(\hat{\vect{p}}\right)\otimes Y_{K_2}\!\left(\hat{\vect{p}}'\right)\right]_{\lambda_0}
 \otimes
 \left[\sigma_{1}(1)\otimes \sigma_{1}(2)\right]_{\lambda_0}
 \!\big]_{00}
$
with $\lambda_0=0$ or $2$.
Therefore, again, the structure of Eq.~\eqref{antisymC6ME} is essentially same as that of other tensor terms,
and differences can be found only in the prefactors and the MPE function $\mathcal{F}_{\lambda_0 K}^{(C_6)}$.
A general form of the tensor MEs is thus obtained in Sec.~\ref{Sec2BMEchiralsummary}.

Note that, following the procedure for the $C_6$-term MEs, 
one can derive the MEs of the $\sigma L$ terms in association with the operator,
$ \left[\vect{\sigma}_1\cdot\left(\vect{q}\times\vect{Q}\right)\right]
 \left[\vect{\sigma}_2\cdot\left(\vect{q}\times\vect{Q}\right)\right]$,
which appears in Eqs.~\eqref{potN3LO2pi} and~\eqref{potN3LOct}.
Indeed, this operator can be written in terms of
$
 \big[\!
 \left[Y_{K_1}\!\left(\hat{\vect{p}}\right)\otimes Y_{K_2}\!\left(\hat{\vect{p}}'\right)\right]_{\lambda_0}
 \otimes
 \left[\sigma_{1}(1)\otimes \sigma_{1}(2)\right]_{\lambda_0}
 \!\big]_{00}
$.
As a result, their MEs have the structure essentially same as that of Eq.~\eqref{antisymC6ME}, i.e., $\lambda_0=0$ and $2$ respectively correspond to the
central and tensor components, although the prefactors and coefficients 
are much more complicated.

\subsubsection{Summary of chiral two-body matrix elements}
\label{Sec2BMEchiralsummary}
In the previous sections, the typical MEs of the central, SO, and tensor contributions are shown.
Now, in this section, we generalize them:
\begin{align}
 V_{ijkl}^{(\nu\nu':C)}
 &=
 A_{ijkl}^{(\nu\nu')}
 \sum_{L=0}^\infty
 U_{ijkl,L}^{(C)}
 P_L(z_{ijkl})
 \nonumber\\
 &\times
 \int\!\!\!\!\int
 dp dp'
 p^{2}p'^{2}
 g_{\nu\nu'}(p,p') u_n\!\left(p,p',\Lambda\right)
 j_{L}\!\left(z_{ij}p\right)  j_{L}\!\left(z_{kl}^*p'\right)
 f_{L}^{(C)}\!\left(p,p'\right),
 \label{genantisym2BMEC}\\
 V_{ijkl}^{(\nu\nu':LS)}
 &=
 A_{ijkl}^{(\nu\nu')}
 \sum_{\lambda_q=0,1}\sum_{\lambda_Q=0,1}
 \sum_{L=0}^\infty \sum_{L'=0}^\infty
 (-)^{\lambda_Q} i^{L+L'}
 \widehat{1\!-\!\lambda_q}\widehat{1\!-\!\lambda_Q}
 \left[\binom{3}{2\lambda_q}\binom{3}{2\lambda_Q}\right]^{\!\frac{1}{2}}
 \nonumber\\
 &\times
 U_{ijkl,L}^{(LS)}
 \left[Y_{L}\!\left(z_{ij,x},z_{ij,y},z_{ij,z}\right) \otimes Y_{L'}\!\left(z_{kl,x}^*,z_{kl,y}^*,z_{kl,z}^*\right)\right]_{1,m_{\sigma_i}+m_{\sigma_j}-m_{\sigma_k}-m_{\sigma_l}}
 \nonumber\\
 &\times
 \sum_{L_1L_2K}
 \left( \lambda_q 0 \lambda_Q 0 | L_1 0 \right)
 \left( 1\!-\!\lambda_q, 0, 1\!-\!\lambda_Q, 0 | L_2 0 \right)
 \nonumber\\
 &\times
 \int\!\!\!\!\int
 dp dp'
 p^{\lambda_q+\lambda_Q+2}p'^{4-\lambda_q-\lambda_Q}
 g_{\nu\nu'}(p,p') u_n\!\left(p,p',\Lambda\right)
 j_{L}\!\left(z_{ij}p\right)  j_{L'}\!\left(z_{kl}^*p'\right)
 f_{LL'L_1L_2K}^{(LS)}\!\left(p,p'\right),
 \label{genantisym2BMELS}\\
 V_{ijkl}^{(\nu\nu':T)}
 &=
 A_{ijkl}^{(\nu\nu')}
 \sum_{\lambda_0=0,2} \sum_{\lambda_1=0}^{\lambda_0}
 \binom{2\lambda_0+1}{2\lambda_1}^{\!\frac{1}{2}}
 \frac{\widehat{\lambda_0\!-\!\lambda_1}}{\hat{\lambda}_0}\left( 1 0 1 0 | \lambda_0 0 \right)
 \nonumber\\
 &\times
 \sum_{L=0}^\infty\sum_{L'=0}^\infty \sum_{K} i^{L+L'}
 \left( \lambda_1 0 K 0 | L 0\right)
 \left( \lambda_0\!-\!\lambda_1, 0 K 0 | L' 0\right)
 \begin{Bmatrix}
  \lambda_0\!-\!\lambda_1 & \lambda_1 & \lambda_0 \\
  L                 & L'       & K
 \end{Bmatrix}
 \nonumber\\
 &\times
 U_{ijkl,L}^{(T)}
 \left[Y_{L}\!\left(z_{ij,x},z_{ij,y},z_{ij,z}\right) \otimes Y_{L'}\!\left(z_{kl,x}^*,z_{kl,y}^*,z_{kl,z}^*\right)\right]_{\lambda_0,m_{\sigma_i}+m_{\sigma_j}-m_{\sigma_k}-m_{\sigma_l}}
 \nonumber\\
 &\times
 \int\!\!\!\!\int
 dp dp'
 p^{\lambda_1+2}p'^{\lambda_0-\lambda_1+2}
 g_{\nu\nu'}(p,p')  u_n\!\left(p,p',\Lambda\right)
 j_{L}\!\left(z_{ij}p\right)  j_{L'}\!\left(z_{kl}^*p'\right)
 f_{\lambda_0 K}^{(T)}\!\left(p,p'\right),
 \label{genantisym2BMET}
\end{align}
\begin{align}
 V_{ijkl}^{(\nu\nu':\sigma L)}
 &=
 A_{ijkl}^{(\nu\nu')}
 \sum_{\lambda_0=0,2}
 \sum_{L_q=0,2} \sum_{L_Q=0,2}
 \left(\frac{1}{2}\right)^{\!L_Q}
 \hat{L}_q \hat{L}_Q
 \left( 1 0 1 0 | L_q 0 \right)
 \left( 1 0 1 0 | L_Q 0 \right)
 \begin{Bmatrix}
  1   & 1   & 1 \\
  1   & 1   & 1 \\
  L_q & L_Q & \lambda_0
 \end{Bmatrix}
 \nonumber\\
 &\times
 \sum_{\lambda_q=0}^{L_q}
 \sum_{\lambda_Q=0}^{L_Q}
 (-)^{\lambda_Q}
 \widehat{L_q\!-\!\lambda_q}\widehat{L_Q\!-\!\lambda_Q}
 \left[\binom{2L_q+1}{2\lambda_q}\binom{2L_Q+1}{2\lambda_Q}\right]^{\!\frac{1}{2}}
 \nonumber\\
 &\times
 \sum_{L_1L_2}
 \hat{L}_1 \hat{L}_2
 \left( \lambda_q 0 \lambda_Q 0 | L_1 0 \right)
 \left( L_q\!-\!\lambda_q, 0, L_Q\!-\!\lambda_Q, 0 | L_2 0 \right)
 \begin{Bmatrix}
  \lambda_q & L_q\!-\!\lambda_q & L_q \\
  \lambda_Q & L_Q\!-\!\lambda_Q & L_Q \\
  L_1       & L_2               & \lambda_0
 \end{Bmatrix}
 \nonumber\\
 &\times
 \sum_{L=0}^\infty \sum_{L'=0}^\infty \sum_{K} i^{L+L'}
 \left( L_1 0 K 0 | L 0\right)
 \left( L_2 0 K 0 | L' 0\right)
 \begin{Bmatrix}
  L   & L' & \lambda_0 \\
  L_2 & L_1 & K
 \end{Bmatrix}
 \nonumber\\
 &\times
 U_{ijkl,L}^{(T)}
 \left[Y_{L}\!\left(z_{ij,x},z_{ij,y},z_{ij,z}\right) \otimes Y_{L'}\!\left(z_{kl,x}^*,z_{kl,y}^*,z_{kl,z}^*\right)\right]_{\lambda_0,m_{\sigma_i}+m_{\sigma_j}-m_{\sigma_k}-m_{\sigma_l}}
 \nonumber\\
 &\times
 \int\!\!\!\!\int
 dp dp'
 p^{\lambda_q+\lambda_Q+2} p'^{\,L_q+L_Q-\lambda_q-\lambda_Q+2}
 \nonumber\\
 &\times
 g_{\nu\nu'}(p,p')  u_n\!\left(p,p',\Lambda\right)
 j_{L}\!\left(z_{ij}p\right)  j_{L'}\!\left(z_{kl}^*p'\right)
 f_{L_qL_QK}^{(\sigma L)}\!\left(p,p'\right).
 \label{genantisym2BMEsigmaL}
\end{align}
When we focus on the central MEs, for example, 
the correspondence between Eqs.~\eqref{antisymC1ME} and~\eqref{genantisym2BMEC}
is clear: By absorbing the prefactors of Eq.~\eqref{antisymC1ME} 
into $\mathcal{F}^{(C_1)}_L$, which is newly defined as $f_{L}^{(C)}$,
and by replacing $D_{ijkl}^{(\sigma)} D_{ijkl}^{(\tau)}-(-)^LD_{jikl}^{(\sigma)} D_{jikl}^{(\tau)}$ with $U_{ijkl,L}^{(C)}$, one can obtain Eq.~\eqref{genantisym2BMEC}.
The explicit form of the MPE functions, 
$f_{L}^{(C)}$, $f_{LL'L_1L_2K}^{(LS)}$, $f_{\lambda_0 K}^{(T)}$, and $f_{L_qL_QK}^{(\sigma L)}$ 
are relegated to Appendix~\ref{SecMPE}.


\begin{table}[!t]
 \begin{center}
  \caption{The spin-isospin MEs $U_{ijkl,L}^{(X)}$ of the chiral interaction. 
  The tensor operator can be $\left(\vect{\sigma}_1\cdot\vect{q}\right)\left(\vect{\sigma}_2\cdot\vect{q}\right)$,
  $\left(\vect{\sigma}_1\cdot\vect{Q}\right)\left(\vect{\sigma}_2\cdot\vect{Q}\right)$,
  and $\left[\vect{\sigma}_1\cdot\left(\vect{q}\times\vect{Q}\right)\right]
 \left[\vect{\sigma}_2\cdot\left(\vect{q}\times\vect{Q}\right)\right]$.
  In the right-most column, the corresponding terms are listed.
  The N$^3$LO potentials are characterized by the superscripts, for which $c_i$, $m_N$, and 2L,
  denote the LEC, the average nucleon mass, and the two-loop contributions, respectively (see Appendix~\ref{SecChiralpotN3LO}).}
  \begin{tabular}{ccl}
   \toprule
   Operator type & $U_{ijkl,L}^{(X)}$ & \multicolumn{1}{c}{Chiral interaction} \\
   \midrule
   \multirow{3}{*}{$\mathbb{1}$} &
       \multirow{3}{*}{$D_{ijkl}^{(\sigma)} D_{ijkl}^{(\tau)}
                       -(-)^LD_{jikl}^{(\sigma)} D_{jikl}^{(\tau)}$} &
	   $C_S$, $C_1$, $C_2$, $V_C^{(3)}$, \\
                               &  &$D_1$, $D_2$, $D_3$, $D_4$, \\
                               &  &$V_C^{(c_i^2)}$, $V_C^{(c_i/m_N)}$, $V_C^{(m_N^{-2})}$, $V_C^{(\mathrm{2L})}$ \\
   \cmidrule(lr){1-3}

   \multirow{2}{*}{$\vect{\sigma}_1\cdot\vect{\sigma}_2$} &
       \multirow{2}{*}{$X_{ijkl}^{(\sigma)} D_{ijkl}^{(\tau)}
                       -(-)^LX_{jikl}^{(\sigma)} D_{jikl}^{(\tau)}$} &
	   $C_T$, $C_3$, $C_4$, $V_S^{(2)}$, $V_S^{(3)}$, \\
                           & & $D_5$, $D_6$, $D_7$, $D_8$, $V_S^{(m_N^{-2})}$, $V_S^{(\mathrm{2L})}$ \\
   \cmidrule(lr){1-3}

   \multirow{2}{*}{$\vect{\tau}_1\cdot\vect{\tau}_2$} &
   \multirow{2}{*}{$D_{ijkl}^{(\sigma)} X_{ijkl}^{(\tau)}
                   -(-)^LD_{jikl}^{(\sigma)} X_{jikl}^{(\tau)}$} &
	   $W_C^{(2)}$, $W_C^{(3)}$, \\
           & &$W_C^{(c_i/m_N)}$, $W_C^{(m_N^{-2})}$, $W_C^{(\mathrm{2L})}$ \\
   \cmidrule(lr){1-3}

   \multirow{2}{*}{$(\vect{\sigma}_1\cdot\vect{\sigma}_2)(\vect{\tau}_2\cdot\vect{\tau}_2)$} &
   \multirow{2}{*}{$X_{ijkl}^{(\sigma)} X_{ijkl}^{(\tau)}
                   -(-)^LX_{jikl}^{(\sigma)} X_{jikl}^{(\tau)}$} &
	   $W_S^{(3)}$, \\
                               & &$W_S^{(c_i^2)}$, $W_S^{(c_i/m_N)}$, $W_S^{(m_N^{-2})}$, $W_S^{(\mathrm{2L})}$ \\
   \cmidrule(lr){1-3}

   \multirow{2}{*}{$-i\vect{S}\cdot\left(\vect{q}\times \vect{Q}\right)$} &
   \multirow{2}{*}{$S_{ijkl} \left[D_{ijkl}^{(\tau)}-(-)^LD_{ijkl}^{(\tau)}\right]$} &
	   $C_5$, $V_{LS}^{(3)}$, \\
                                 & &$D_9$, $D_{10}$, $V_{LS}^{(c_i/m_N)}$, $V_{LS}^{(m_N^{-2})}$ \\
   \cmidrule(lr){1-3}

   $-i\vect{S}\cdot\left(\vect{q}\times \vect{Q}\right)$ & \multirow{2}{*}{$S_{ijkl} \left[X_{ijkl}^{(\tau)}-(-)^LX_{jikl}^{(\tau)}\right]$} &
   \multirow{2}{*}{$W_{LS}^{(3)}$, $W_{LS}^{(c_i/m_N)}$, $W_{LS}^{(m_N^{-2})}$}  \\
   $\times\vect{\tau}_1 \cdot \vect{\tau}_2$ & &\\
   \cmidrule(lr){1-3}

   \multirow{3}{*}{Tensor} & \multirow{3}{*}{$T_{ijkl}^{(\lambda_0)}D_{ijkl}^{(\tau)}-(-)^LT_{jikl}^{(\lambda_0)}D_{jikl}^{(\tau)}$} &  $C_6$, $C_7$, $V_T^{(2)}$, $V_T^{(3)}$, \\
            & & $D_{11}$, $D_{12}$, $D_{13}$, $D_{14}$, $D_{15}$, \\
                               &  &$V_T^{(m_N^{-2})}$, $V_{\sigma L}^{(m_N^{-2})}$, $V_T^{(\mathrm{2L})}$ \\
   \cmidrule(lr){1-3}

   \multirow{2}{*}{(Tensor)$\vect{\tau}_1\cdot\vect{\tau}_2$} &
       \multirow{2}{*}{$T_{ijkl}^{(\lambda_0)}X_{ijkl}^{(\tau)}-(-)^LT_{jikl}^{(\lambda_0)}X_{jikl}^{(\tau)}$} &
	   1$\pi$, $W_T^{(3)}$, \\
   	   & &$W_T^{(c_i^2)}$, $W_T^{(c_i/m_N)}$, $W_T^{(m_N^{-2})}$, $W_T^{(\mathrm{2L})}$\\
   \bottomrule
  \end{tabular}
  \label{tableChiralSpinIso}
 \end{center}
\end{table}
Each term of the MEs involves $U_{ijkl,L}^{(X)}$, which depends on the operator form
as summarized in Table~\ref{tableChiralSpinIso}.
For instance, we explicitly show the correspondence between $U_{ijkl,L}^{(X)}$ 
and the spin-isospin MEs of the $C_1$, $C_5$, and $C_6$ terms derived in the previous sections.
They are characterized by the operators, $\mathbb{1}$, $-i\vect{S}\cdot\left(\vect{q}\times \vect{Q}\right)$,
and $\left(\vect{\sigma}_1\cdot\vect{q}\right)\left(\vect{\sigma}_2\cdot\vect{q}\right)$:
\begin{align}
 U_{ijkl,L}^{(C)}
 &=
 D_{ijkl}^{(\sigma)} D_{ijkl}^{(\tau)}
 -(-)^L D_{jikl}^{(\sigma)} D_{jikl}^{(\tau)},
 \qquad\left[\mathrm{for}~\mathbb{1}\right],
 \label{DDDD}\\
 U_{ijkl,L}^{(LS)}
 &=
 S_{ijkl}
 \left[D_{ijkl}^{(\tau)}-(-)^L D_{jikl}^{(\tau)}\right]
 \qquad\left[\mathrm{for}~-i\vect{S}\cdot\left(\vect{q}\times \vect{Q}\right)\right],
 \label{SDD}\\
 U_{ijkl,L}^{(T)}
 &=
 T_{ijkl}^{(\lambda_0)}D_{ijkl}^{(\tau)}-(-)^LT_{jikl}^{(\lambda_0)}D_{jikl}^{(\tau)}
 \quad\left[\mathrm{for}~\left(\vect{\sigma}_1\cdot\vect{q}\right)\left(\vect{\sigma}_2\cdot\vect{q}\right)\right].
 \label{TDD}
\end{align}
If the interaction involves the spin-spin operator, $\vect{\sigma}_1 \cdot \vect{\sigma}_2$,
we have to replace $D_{ijkl}^{(\sigma)}$ and $D_{jikl}^{(\sigma)}$ respectively with 
$X_{ijkl}^{(\sigma)}$ and $X_{jikl}^{(\sigma)}$ defined by
\begin{align}
 X_{ijkl}^{(\sigma)}
 &=
 2\delta_{m_{\sigma_i} m_{\sigma_l}} \delta_{m_{\sigma_j} m_{\sigma_k}}
 -\delta_{m_{\sigma_i} m_{\sigma_k}} \delta_{m_{\sigma_j} m_{\sigma_l}}.
 \label{Xijkl}
\end{align}
The same is true for the isospin MEs, i.e., $D_{ijkl}^{(\tau)}$ and $D_{jikl}^{(\tau)}$ 
respectively become $X_{ijkl}^{(\tau)}$ and $X_{jikl}^{(\tau)}$ with the operator $\vect{\tau}_1 \cdot \vect{\tau}_2$.
Here, $\vect{\tau}_i$ is the isospin operator represented by the Pauli matrices.
The isospin-isospin operator, $\vect{\tau}_1 \cdot \vect{\tau}_2$,
does not enter the contact terms of the usual chiral EFT
since they are formulated with the choice of $\mathbb{1}$ and $\vect{\sigma}_1 \cdot \vect{\sigma}_2$
based on the Fierz rearrangement freedom~\cite{PhysRevLett.116.062501,PhysRevC.96.054003,10.3389/fphy.2019.00245}.


\section{Visualization of two-body matrix elements}
\label{SecVis}
\subsection{Numerical details}
\label{SecVisDetail}
\begin{figure}[!b]
\begin{center}
\includegraphics[width=0.4\textwidth,clip]{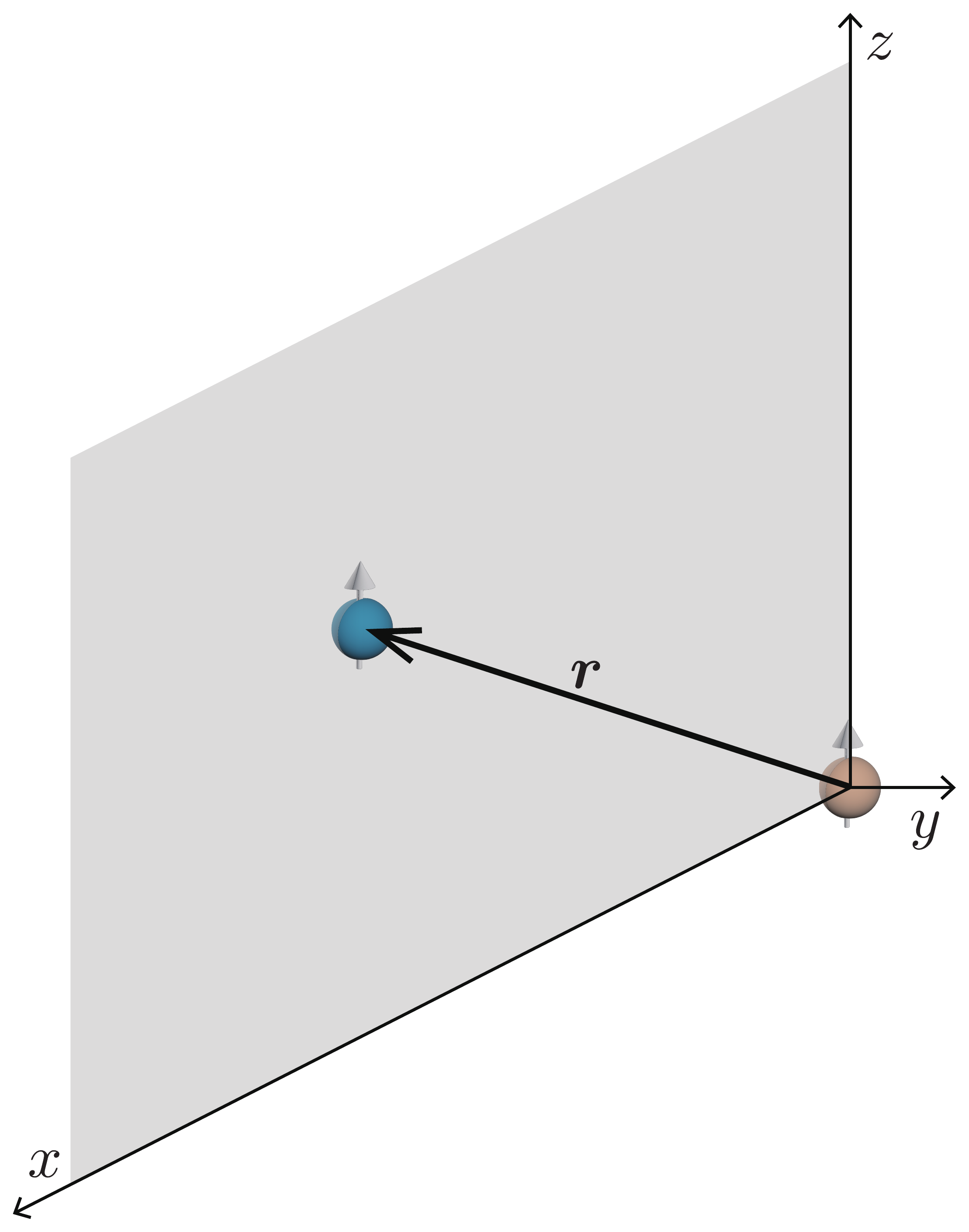}
 \caption{The geometric configuration for the visualization of the two-body MEs.
 A spin-up neutron (spin-up proton) is settled at the origin (on the $x$-$z$ plane).}
\label{figintuitive}
\end{center}
\end{figure}
To demonstrate the behavior of the two-body MEs formulated in Sec.~\ref{SecMEs},
here we visualize typical values of them.
As an example, we choose the LO (LO plus NLO) contributions, 
the potentials of which are given in Appendix~\ref{SecChiralpotLO} (Appendix~\ref{SecChiralpotNLO}).
To visualize the MEs some constrains are necessary.
First, the Gaussian-center position $\vect{Z}_i$ is chosen to be real, 
and we select the diagonal MEs $V_{ijij}^{(\nu\nu)}$.
Thus, the MEs can be computed as a function of $\vect{r}=\vect{z}_{ij}=\vect{z}_{kl}$,
which is the relative distance between two nucleons.
Next, as depicted in Fig.~\ref{figintuitive}, a spin-up neutron is settled at the origin, $\vect{r}=0$, and another spin-up proton moves on the $x$-$z$ plane,
i.e., now the three-dimensional vector $\vect{r}$ is expressed by $\vect{r}=(x,0,z)$.
Note that the spin direction is aligned with the $z$ axis, and therefore, 
the effect of the tensor contributions originating from the $1\pi$ exchange at LO, 
as well as from $2\pi$ exchange and contacts at NLO,
can be seen on the $x$-$z$ plane.
Hereafter we use the shorthand notations, $\Braket{\hat{V}_{\mathrm{LO}}}$
and $\Braket{\hat{V}_{\mathrm{NLO}}}$, 
to express the MEs $V_{ijij}^{(\nu\nu)}$ of the LO and LO-plus-NLO cases, respectively.

The parameters we employ here are summarized in Ref.~\cite{Fukui_2022}, 
where the LECs are originally taken from~\cite{MACHLEIDT20111,PhysRevC.87.014322,PhysRevC.89.044321}.
One can also find the regulator parameters, $\Lambda$ and $n$,
as well as the constants relevant to the $1\pi$ term (the pion mass, pion-decay constant, and axial vector coupling constant).
We adopt $\nu=0.26$~fm$^{-2}$~\cite{Fukui_2022}.

\subsection{Matrix elements at leading order and contact-term contributions}
\label{SecVisFullLO}
\begin{figure}[H]
\begin{center}
\includegraphics[width=1.0\textwidth,clip]{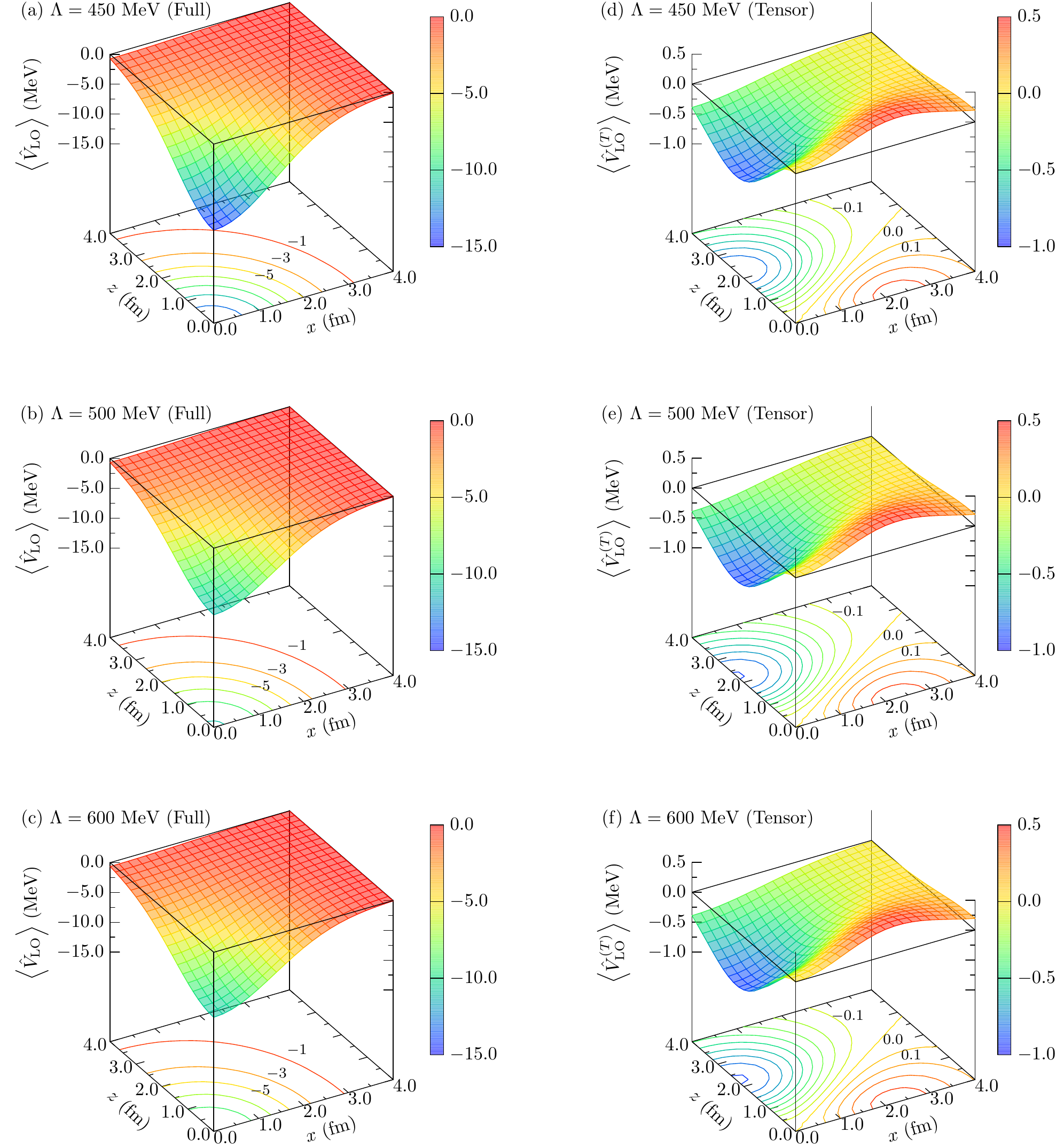}
 \caption{The two-body MEs of the chiral interaction at LO 
with the cutoff (a) $\Lambda=450$, (b) $500$, and (c) $600$~MeV, with $x$ and $z$ corresponding to the axes shown in Fig.~\ref{figintuitive}.
The MEs of the OPEP-tensor contributions are also displayed with (d) $\Lambda=450$, (e) $500$, and (f) $600$~MeV.
See the text for detail.}
\label{figallinoneLO}
\end{center}
\end{figure}

Under the conditions described in the previous section, $\Braket{\hat{V}_{\mathrm{LO}}}$ can be represented in Figs.~\ref{figallinoneLO}(a)-(c).
The results with three different values of $\Lambda$ are shown ($\Lambda=450$, $500$, and $600$~MeV).
One finds that the larger $\Lambda$ is, the less attractive $\Braket{\hat{V}_{\mathrm{LO}}}$ is. 
This is due to the $\Lambda$ dependence of the LECs at LO,
i.e., $C_S$ ($C_T$), which is responsible for the attraction (repulsion),
becomes smaller (larger) as $\Lambda$ increases~\cite{MACHLEIDT20111,PhysRevC.87.014322,PhysRevC.89.044321}.

Figures~\ref{figallinoneLO}(a)-(c) show that the spin-aligned-neutron-proton pair feels the largest attraction at $\vect{r}=0$.
This is because the LO potential is designed to simulate nucleon-nucleon scattering at very low momentum,
where the neutron-proton interaction of the triplet-$s$ state is attractive as deduced from the scattering phase shift~\cite{PhysRevC.48.792}.
Furthremore, the attractive MEs at $\vect{r}=0$ can be shown analytically.
First, we ignore the regulator with $\Lambda\to\infty$.
Indeed, the largest attraction at $\vect{r}=0$ can be seen independently of $\Lambda$, 
and therefore, the regulator does not play an essential role for the present discussion.
Then, the MEs of the LO-contact term 
within the configuration of Fig.~\ref{figintuitive} can be simplified as
\begin{align}
 \Braket{\hat{V}_{\textrm{LO-ct}}}
 \xrightarrow{\Lambda\to\infty}
 \left(\frac{\nu}{\pi}\right)^{\!\frac{3}{2}}
 (C_S^{np}+C_T^{np})\exp\!\left[-\nu r^2\right].
 \label{LOctMEsspecial}
 \end{align}
Note that the LECs at LO has the charge dependence.
The superscript $np$ stands for the LECs for the neutron-proton pair.
We find that the condition $C_S^{np}+C_T^{np} < 0$ is satisfied by the LECs employed here.
For example, $C_S^{np}+C_T^{np} =-0.011374\times 10^4$~GeV$^{-2}$ for $\Lambda=450$~MeV~\cite{PhysRevC.87.014322,PhysRevC.89.044321}.
Thus we can show that Eq.~\eqref{LOctMEsspecial} has a minimum at $\vect{r}=0$.

We should mention a role played by the $1\pi$ interaction at the origin.
Within the configuration of Fig.~\ref{figintuitive} 
at $\vect{r}=0$, the system is a triplet-even state, and hence,
the central term of the $1\pi$-exchange interaction is repulsive, 
while the $1\pi$-tensor term has no contributions there [see Figs.~\ref{figallinoneLO}(d)-(f)].

Note that the $1\pi$-exchange potential (OPEP) considered here contains 
the short-range delta function in coordinate space.
This short-range term is often subtracted by hand since $1\pi$ exchange should be responsible for 
the long-range part of the two-nucleon force (see Ref.~\cite{Yamaguchi_2020} for example),
and thus, the central term of this ``subtracted'' OPEP is attractive for the triplet-even state.
Instead, in the chiral EFT, the short-range term of the OPEP remains included
but the addition of the LO-contact term as a counterterm suppresses the short-range repulsion of the OPEP.
Consequently, for the triplet-even state, the OPEP-central term of the chiral EFT alone is repulsive,
but the whole LO potential is attractive.

The repulsion by the $1\pi$-contact term is moderate compared to that by the LO-contact terms.
Specifically, for $\Lambda=450$~MeV, the absolute vale of the $1\pi$-central ME at the origin is less than $30\%$ of that of the contacts.
Therefore, as addressed above, the minimum of the MEs at $\vect{r}=0$ 
can be basically explained by the contributions from the LO-contact terms.

\subsection{One-pion-exchange-tensor contributions to matrix elements}
\label{SecVisTensorLO}
By carefully watching Fig.~\ref{figallinoneLO}(a)-(c), one finds that the MEs are not symmetric with respect to the $z=x$ line.
For example, the contour line of $-1$~MeV in Fig.~\ref{figallinoneLO}(a) crosses the $z$ and $x$ axes at $\sim 4$
and $\sim 3$ fm, respectively. Hence, the MEs are more attractive in the $z>x$ region and vice versa.
This asymmetry is due to the tensor contributions of the $1\pi$-exchange term.

Analytically, the asymmetry can be understood as follows.
The tensor operator defined by Eq.~\eqref{tensope1} has anisotropy.
As well known (see Ref.~\cite{Yamaguchi_2020} for example), the OPEP in the coordinate-space representation,
which is the Fourier transform of Eq.~\eqref{potLO1pi},
involves the tensor operator,
\begin{align}
 \mathcal{S}_{12}(\hat{\vect{r}})
 &=
 2\left[3\left(\vect{S}\cdot\hat{\vect{r}}\right)^2
 -\vect{S}^2\right].
 \label{tensope2}
\end{align}
The direction of $\vect{S}$ is now aligned with the $z$ axis,
and therefore, the attraction by the tensor contributions becomes stronger for the $z>x$ region on the $x$-$z$ plane.
Note that the expectation value of the isospin operator $\vect{\tau}_1\cdot\vect{\tau}_2$ appearing in the OPEP
is $-3$ for the isoscalar state at $\vect{r}=0$. 
For finite $\vect{r}$, the isoscalar-isovector mixing occurs and it is taken into account in the present calculations.

Moreover, one can realize that the bipolar spherical harmonics, which is involved in the tensor MEs given by Eq.~\eqref{genantisym2BMET},
is asymmetric with respect to the exchange of $x$ and $z$.
Indeed, under the present conditions, the bipolar spherical harmonics of the tensor component ($\lambda_0=2$) becomes
\begin{align}
 &\left[Y_{L}\!\left(z_{ij,x},z_{ij,y},z_{ij,z}\right) \otimes Y_{L'}\!\left(z_{kl,x}^*,z_{kl,y}^*,z_{kl,z}^*\right)\right]_{\lambda_0,m_{\sigma_i}+m_{\sigma_j}-m_{\sigma_k}-m_{\sigma_l}}
 \nonumber\\
 &\quad\to
 \frac{1}{8\pi}\hat{L}\hat{L}'\left(L0L'0|20\right)
 \frac{2z^2-x^2}{x^2+z^2},
 \label{BipolarYYtensor}
\end{align}
for which we notice the asymmetry with respect to the operation $x\leftrightarrow z$.

Numerically, the asymmetry of the tensor MEs $\Braket{\hat{V}_{\mathrm{LO}}^{(T)}}$ can be clearly seen in Figs.~\ref{figallinoneLO}(d)-(f),
where only the tensor component $\lambda_0=2$ of Eq.~\eqref{genantisym2BMET} for the OPEP is taken into account.
Figure~\ref{figallinoneLO}(d)-(f) displays that the tensor contributions are attractive (repulsive) on the $z$ ($x$) axis,
as expected from the structure of the tensor operator, Eq.~\eqref{tensope2}.

At $\vect{r}=0$, even though the radial part of the tensor component of the OPEP diverges 
(see Ref.~\cite{Machleidt1989} for example), 
the tensor MEs must be zero, since only the $s$ wave contributes when two nucleons contact with each other.
Note that the divergence of the OPEP-tensor part does not matter in the present calculations
since the MEs are computed in momentum space, where the corresponding high-momentum component of the OPEP is suppressed by the regulator.
The position of the extrema of $\Braket{\hat{V}_{\mathrm{LO}}^{(T)}}$ depends on $\nu$.
If we compute the tensor MEs with larger $\nu$, 
the minimum (maximum) point moves in the direction that $z$ ($x$) becomes smaller, 
and the attractive pockets in Figs.~\ref{figallinoneLO}(d)-(f) become deeper with larger $\nu$.
This is consistent with the $\nu$ dependence of tensor-force contributions to
the $^4\mathrm{He}$ energy reported in Ref.~\cite{10.1093/ptep/ptx089};
the energy gain by a tensor force increases when $\nu$ is greater than 
its typical value $\nu=0.25$~fm$^{-2}$ for $^4\mathrm{He}$~\cite{10.1093/ptep/ptx143,PhysRevC.97.014304},
although the total energy is saturated due to the compensation by the energy loss of the kinetic term for a such large $\nu$.

Also one finds from Figs.~\ref{figallinoneLO}(d)-(f) that both attraction and repulsion are enhanced as $\Lambda$ increases.
This results from the operator form of the tensor potential.
In momentum space, the tensor component of the OPEP can be written as
\begin{align}
 \mathcal{O}_{12}(\vect{q})
 =
 \frac{1}{3\left\{1+(m_\pi/q)^2\right\}}\mathcal{S}_{12}(\hat{\boldsymbol{q}}),
 \label{tensorOPEPmom}
\end{align}
which can be obtained from the operator,
$\left(\boldsymbol{\sigma}_1\cdot\boldsymbol{q}\right)\left(\boldsymbol{\sigma}_2\cdot\boldsymbol{q}\right)\left(q^2+m_\pi^2\right)^{-1}$
through the irreducible tensor representation as Eq.~\eqref{sigqsigqC6} with $\lambda_0=2$.
When large $q$ contributes (this is the case for larger $\Lambda$), the effect of $\mathcal{O}_{12}$ is enhanced.
This is consistent with the $\Lambda$ dependence of $\Braket{\hat{V}_{\mathrm{LO}}^{(T)}}$.

\subsection{Matrix elements at next-to-leading order}
\label{SecVisNLO}
\begin{figure}[!t]
\begin{center}
\includegraphics[width=1.0\textwidth,clip]{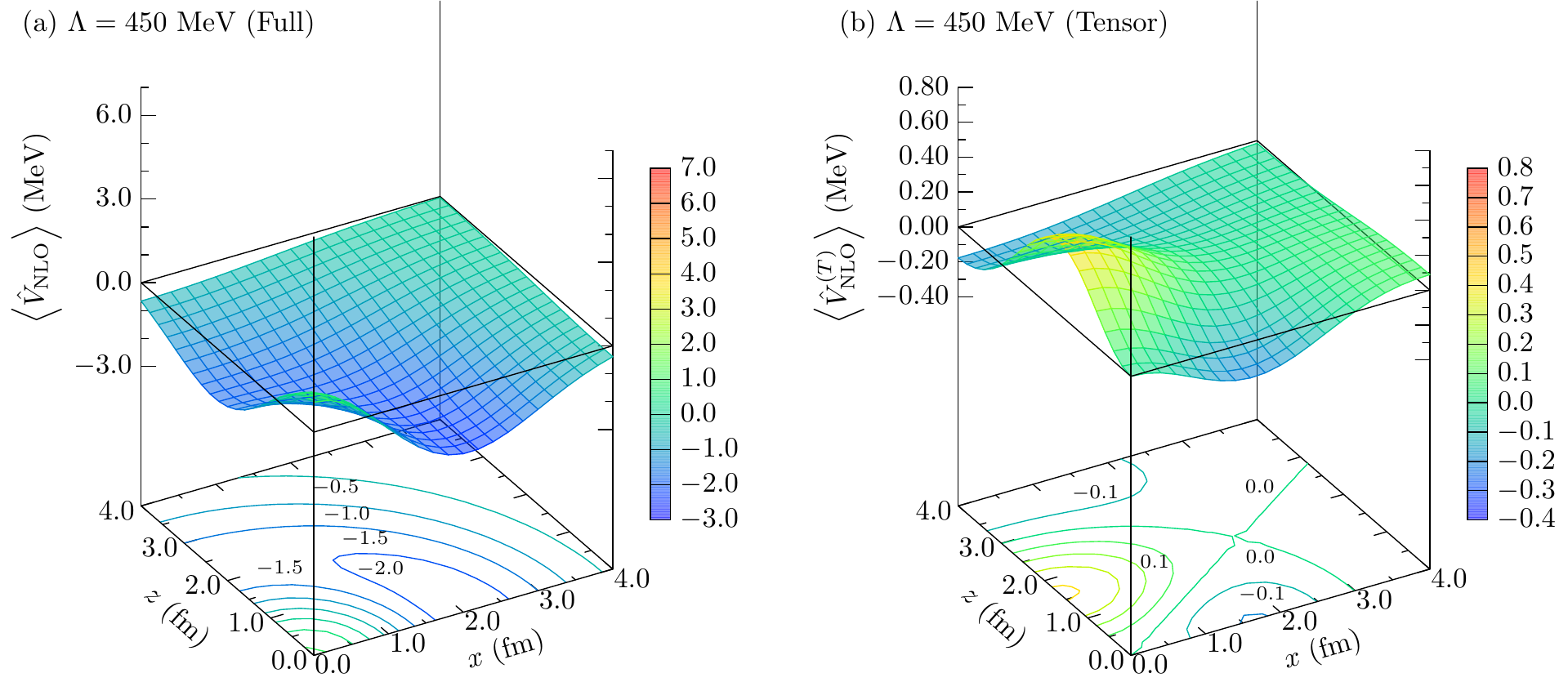}
 \caption{The two-body MEs of the chiral nucleon-nucleon force at NLO for 
 (a) the full contributions and (b) tensor contributions only,
 where $x$ and $z$ axes are given in Fig.~\ref{figintuitive}. 
 The regulator cutoff is $\Lambda=450$ MeV.}
\label{figNLO450}
\end{center}
\end{figure}
\begin{figure}[!t]
\begin{center}
\includegraphics[width=1.0\textwidth,clip]{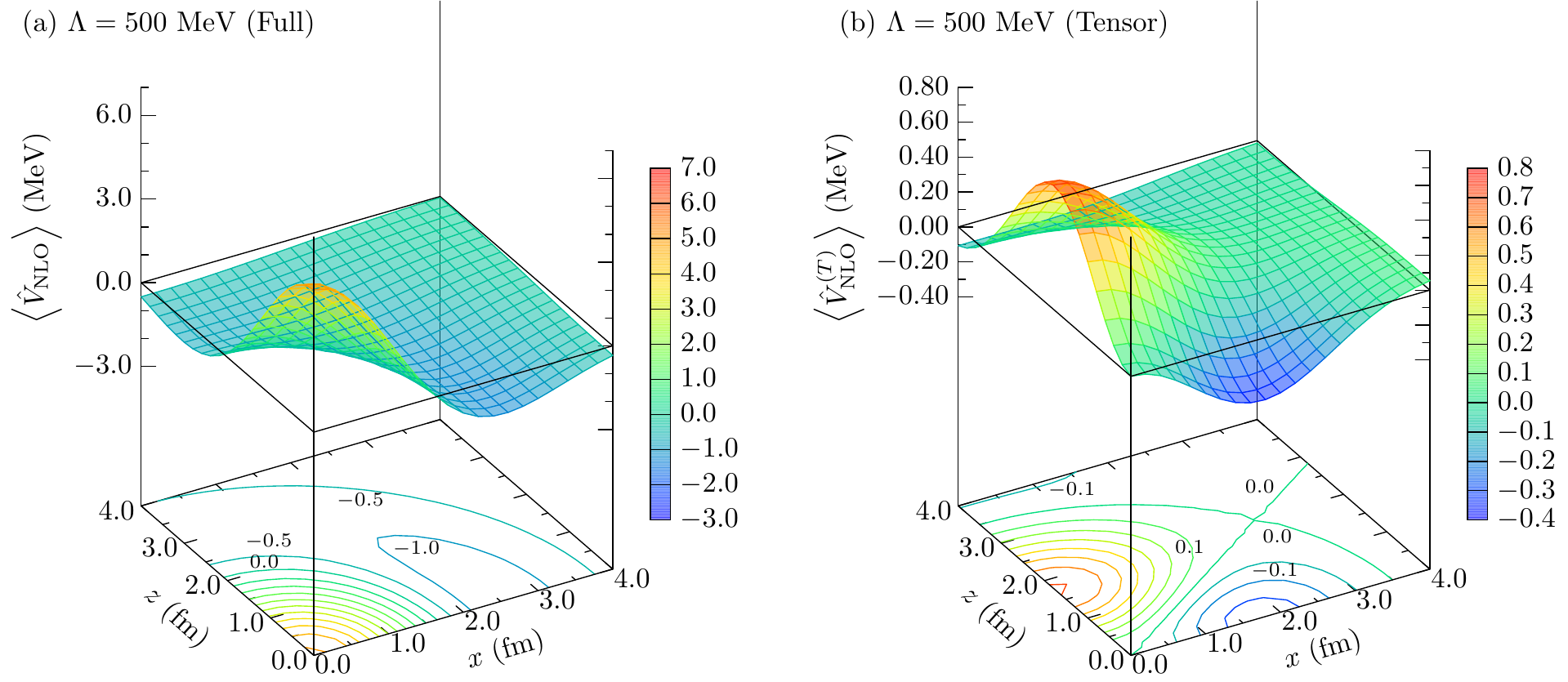}
 \caption{Same as Fig.~\ref{figNLO450} but for $\Lambda=500$ MeV.}
\label{figNLO500}
\end{center}
\end{figure}
\begin{figure}[!t]
\begin{center}
\includegraphics[width=1.0\textwidth,clip]{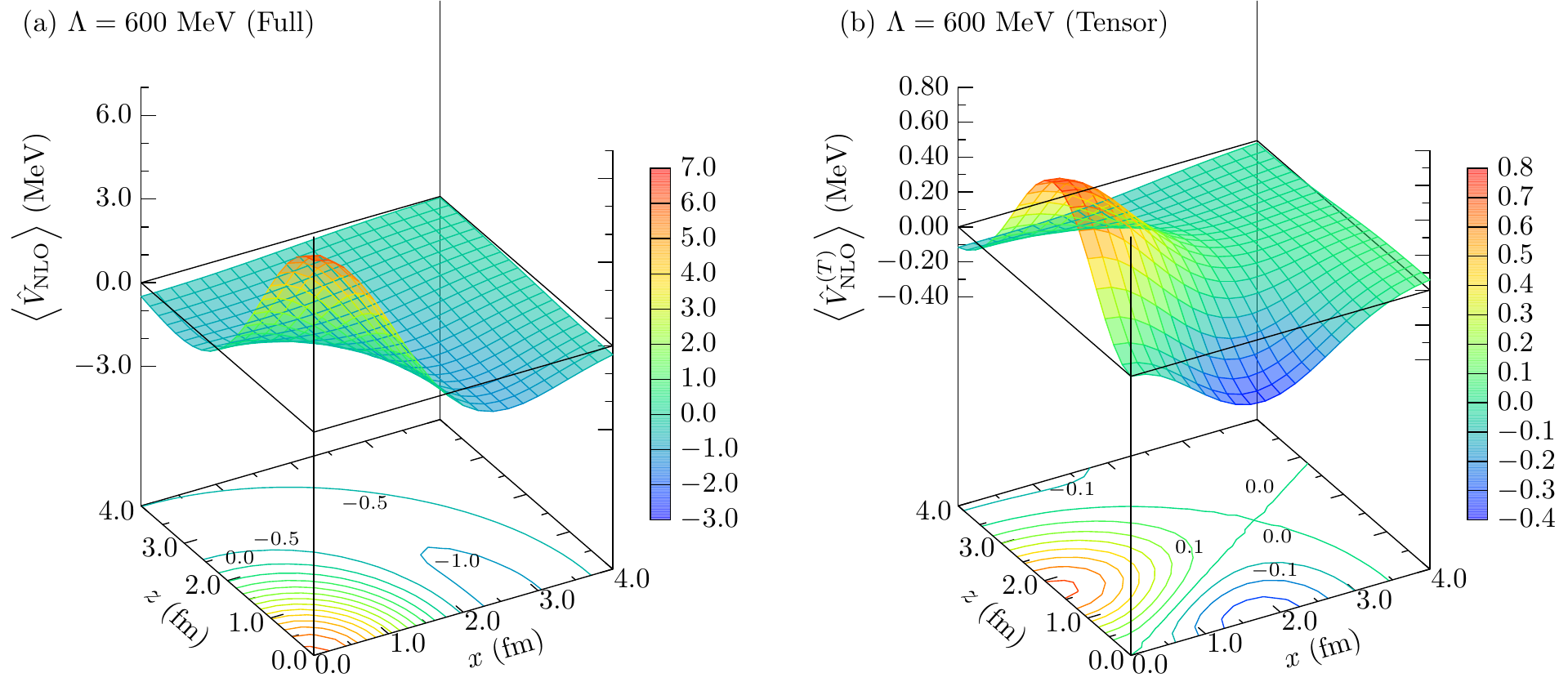}
 \caption{Same as Fig.~\ref{figNLO450} but for $\Lambda=600$ MeV.}
\label{figNLO600}
\end{center}
\end{figure}
Now we increase the order of the chiral EFT up to NLO
to visualize the two-body MEs, $\Braket{\hat{V}_{\mathrm{NLO}}}$, within the two-nucleon configuration same as that in the previous section.
Figure~\ref{figNLO450}(a) shows $\Braket{\hat{V}_{\mathrm{NLO}}}$ 
for the regulator cutoff $\Lambda=450$ MeV, where $x$ and $z$ axes are given in Fig.~\ref{figintuitive}.
The NLO MEs of the tensor contributions denoted by $\Braket{\hat{V}_{\mathrm{NLO}}^{(T)}}$ are extracted from $\Braket{\hat{V}_{\mathrm{NLO}}}$,
and depicted in Fig.~\ref{figNLO450}(b).
The similar results for $\Lambda=500$ and $600$ MeV are shown in Figs.~\ref{figNLO500} and~\ref{figNLO600}, respectively.
Note that even though the $C_5$ term, one of the NLO contact terms, enters the NLO potential as the SO contributions,
it plays no roles on the MEs visualized in Figs.~\ref{figNLO450} to~\ref{figNLO600}.
This is because the direction of $\vect{S}$ is aligned with the $z$ axis and also the Gaussian center is real.
Indeed one can show analytically that the bipolar spherical harmonics involved in Eq.~\eqref{antisymC5ME} becomes zero in the $x$-$z$ plane.

As argued in Sec.~\ref{SecVisFullLO}, the LO MEs at $\vect{r}=0$ are attractive
since the LO potential is tailored to very low-momentum scattering of two nucleons.
In contrast to the LO MEs, the NLO MEs in Figs.~\ref{figNLO450}(a),~\ref{figNLO500}(a), and~\ref{figNLO600}(a)
are all repulsive at $\vect{r}=0$.
This repulsive nature stems from the high-momentum scattering described by the NLO potentials
dependent on the square of the momentum (see Appendix~\ref{SecChiralpotNLO}).
Furthermore, one finds that the more $\Lambda$ increases, the more the repulsion of the MEs enhances.
This $\Lambda$ dependence of the MEs reflects the property of the neutron-proton interaction,
which turns from attractive to repulsive around $400$~MeV$/c$ of the relative momentum 
as indicated by the neutron-proton scattering phase shift of the triplet-$s$ state~\cite{PhysRevC.48.792}.

In Figs.~\ref{figNLO450}(b),~\ref{figNLO500}(b), and~\ref{figNLO600}(b),
the tensor MEs of the LO-plus-NLO contributions, $\Braket{\hat{V}_{\mathrm{NLO}}^{(T)}}$,
clearly show the asymmetry with respect to the $z=x$ line.
The origin of this asymmetry is same as that in the OPEP-tensor MEs, as already discussed in Sec.~\ref{SecVisTensorLO}.
However, $\Braket{\hat{V}_{\mathrm{NLO}}^{(T)}}$ has the opposite sign compared to that of $\Braket{\hat{V}_{\mathrm{LO}}^{(T)}}$,
i.e., the repulsive peak (attraction pocket) of $\Braket{\hat{V}_{\mathrm{NLO}}^{(T)}}$ localizes on the $z$ ($x$) axis,
and vice versa for $\Braket{\hat{V}_{\mathrm{LO}}^{(T)}}$.
This is because the NLO potentials do not depend on the isospin-isospin operator, $\vect{\tau}_1\cdot\vect{\tau}_2$,
which induces the change of the sign of the MEs,
and also because the LECs, $C_6$ and $C_7$ are negative in the present parameterization~\cite{MACHLEIDT20111,PhysRevC.87.014322,PhysRevC.89.044321}.


\section{Conclusions and perspectives}
\label{SecConc}
We have presented the formalism of the two-body MEs of the two-nucleon forces in momentum space derived from the chiral EFT
based on the single-nucleon wave functions expressed by the Gaussian-wave packet.
Such MEs are relevant to many-body calculations like AMD, which can be applicable, for instance, to efficiently describe 
nuclear cluster structures.
We adopt the chiral potentials at N$^3$LO based on the nonlocal regularization.

We have visualized the MEs formulated in this paper by selecting the spin-up neutron and spin-up proton pair,
for which the tensor contributions can be seen.
As an example, the chiral potentials at LO have been chosen.
We have addressed the cutoff dependence of the MEs and the origin of the ME extrema, 
as well as the individual contributions of the central and tensor forces.

As a next step of this work, we are now implementing this formalism to AMD.
Then, benchmark calculations will be performed for light nuclei.
The inclusion of the chiral three-nucleon force into AMD is one of the important tasks,
and the extension of the framework to this direction is also on going.


\section*{Acknowledgment}
The author thanks N. Itagaki and M. Kimura for fruitful discussions and useful comments.
This work was supported by Japan Society for the Promotion of Science KAKENHI with Grant Number JP21K13919.
The calculations were carried out using the computer facilities 
at Yukawa Institute for Theoretical Physics, Kyoto University.

\appendix
\section{Slater determinants and their superposition}
\label{SecMBWFProj}
\subsection{Slater determinants}
\label{SecMBWFProjSlater}
A single-Slater determinant for a system of the mass number $A$ is characterized by a single value of $\nu$
and a set of the generator coordinate $\mathcal{Z}_n$:
\begin{align}
 \Ket{\Psi_\nu(\mathcal{Z}_n)} &
 = \sqrt{A!}\hat{\mathcal{A}}_{A}
 \Ket{\varphi_1^{(\nu)} \varphi_2^{(\nu)} \cdots \varphi_i^{(\nu)} \cdots \varphi_A^{(\nu)}},
 \label{mbwfnu}\\
 \mathcal{Z}_n&=\left\{\vect{Z}_1,\vect{Z}_{2},\cdots,\vect{Z}_{A}\right\}_n,
 \label{RepresentativeZ}
\end{align}
with the $A$-body antisymmetrizer $\hat{\mathcal{A}}_{A}$.
Then, many-body states are obtained through the generator-coordinate method (GCM)~\cite{brink1966proc,PhysRev.89.1102,PhysRev.108.311} after the parity and angular-momentum projections:
\begin{align}
 \Ket{\Psi_{\mathrm{GCM}}}
 &=
 \sum_{n\nu}c_n^{(\nu)}
 \Ket{\Psi_{\nu MK}^{J\pm}(\mathcal{Z}_n)},
 \label{GCMwfgeneral}\\
 \Ket{\Psi_{\nu MK}^{J\pm}(\mathcal{Z}_n)}
 &=
 \hat{\mathcal{P}}_{MK}^J \hat{\mathcal{P}}^{\pm} \Ket{\Psi_{\nu}(\mathcal{Z}_n)},
 \label{Jpiprojectedgeneral}
\end{align}
where the projection operators $\hat{\mathcal{P}}^{\pm}$ and $\hat{\mathcal{P}}_{MK}^J$
are given explicitly in the next sections.

The coefficient $c_n^{(\nu)}$ is obtained numerically by solving the generalized eigenvalue problem,
\begin{align}
 \sum_{m\nu'}\left(H_{nm}^{(\nu\nu')}-EN_{nm}^{(\nu\nu')}\right)c_m^{(\nu')}=0,
 \label{HWeq}
\end{align}
with the energy eigenvalue $E$.
Since the single-particle wave function expressed by the Gaussian-wave packet,
${\Ket{\Psi_\nu(\mathcal{Z}_n)}}$ can be separated into the CM and intrinsic wave functions.
Thus, the MEs of the norm and Hamiltonian $\hat H$ relevant to the intrinsic structure are given by
\begin{align}
 \begin{Bmatrix}
  N_{nm}^{(\nu\nu')}\\
  H_{nm}^{(\nu\nu')}
 \end{Bmatrix}
 &=
 \frac{
 \Braket{\Psi_{\nu MK}^{J\pm}(\mathcal{Z}_n) \left|
 \begin{Bmatrix}
  \mathbb{1}\\
  \hat H
 \end{Bmatrix}
 \right| \Psi_{\nu' MK}^{J\pm}(\mathcal{Z}_m) }
 }
 {\Braket{\Psi_\nu^{(\mathrm{CM})}|\Psi_{\nu'}^{(\mathrm{CM})}}},
 \label{IntrinsicExpVal}
\end{align}
where the CM-wave function $\Ket{\Psi_\nu^{(\mathrm{CM})}}$ is given by
\begin{align}
 \Psi_\nu^{(\mathrm{CM})}\!\left(\vect{r}_{\mathrm{G}}-\vect{Z}_{\mathrm{G}}\right)
 &=
 \left(\frac{2\nu}{\pi}\right)^{\!\frac{3}{4}}
 \exp\!\left[-A\nu\left(\vect{r}_{\mathrm{G}}-\vect{Z}_{\mathrm{G}}\right)^2\right],
 \label{CMSlater}
\end{align}
with $\vect{r}_{\mathrm{G}}=\sum_{i=1}^A \vect{r}_i/A$ and $\vect{Z}_{\mathrm{G}}=\sum_{i=1}^A \vect{Z}_i/A$.

The Hamiltonian $\hat H$ is written as
\begin{align}
 \hat{H}=\hat{T}-\hat{T}_\mathrm{G}+\hat{V}
 \label{Hamilgeneral}
\end{align}
where $\hat{T}$ is the kinetic-energy operator expressed by the sum of the one-body kinetic-energy operator
and $\hat{T}_\mathrm{G}$ is the kinetic-energy operator of the CM system.
The expectation values of these operators can be calculated analytically.
For example, as shown in Ref.~\cite{PhysRevC.106.044310}, one finds
\begin{align}
 \frac{\Braket{\Psi_\nu(\mathcal{Z}_n)\left|\hat{T}_\mathrm{G}\right|\Psi_{\nu'}(\mathcal{Z}_m)}}{\Braket{\Psi_\nu(\mathcal{Z}_n)|\Psi_{\nu'}(\mathcal{Z}_m)}}
 =
 \frac{3\hbar^2\nu\nu'}{m_N(\nu+\nu')}.
 \label{CMkinetic}
\end{align}
Note that Eq.~\eqref{CMkinetic} is obtained under the assumption, $\vect{Z}_{\mathrm{G}}=0$, for both bra and ket states.
The interaction operator $\hat{V}$ consists of nuclear and Coulomb parts,
and if it involves the chiral two-nucleon force,
$H_{nm}^{(\nu\nu')}$ can be written in terms of the two-body MEs given in Sec.~\ref{SecMEs}.

\subsection{Parity projection}
\label{SecCompProjectionParity}
First, we introduce the parity inversion operator $\hat{\mathcal{P}}_\pi$,
which inverts the sign of the Gaussian center position of the single-particle states as
\begin{align}
 \hat{\mathcal{P}}_\pi \Ket{\Psi_\nu(\mathcal{Z})}
 &=\Ket{\Psi_\nu(-\mathcal{Z})},
 \label{ParityInversionMBWFcomplex}\\
 -\mathcal{Z}
 &=\left\{-\vect{Z}_1,-\vect{Z}_{2},\cdots,-\vect{Z}_{A}\right\}.
 \label{Representative-R}
\end{align}
Note that the subscript $n$ is omitted from $\mathcal{Z}$ for simplicity.
Thus, the parity projected operator and parity projected states are respectively defined by
\begin{align}
 \hat{\mathcal{P}}^{\pm}
 &=
 \frac{1\pm \hat{\mathcal{P}}_\pi}{2\alpha_\pm},
 \label{ParityProjOpeRecall}\\
 \Ket{\Psi_\nu^{\pm}}
 &=\hat{\mathcal{P}}^{\pm} \Ket{\Psi_\nu(\mathcal{Z})}
 =
 \begin{cases}
  \frac{1}{2\alpha_+}\left[\Ket{\Psi_\nu(\mathcal{Z})}+\Ket{\Psi_\nu(-\mathcal{Z})}\right], \\
  \frac{1}{2\alpha_-}\left[\Ket{\Psi_\nu(\mathcal{Z})}-\Ket{\Psi_\nu(-\mathcal{Z})}\right],
 \end{cases}
 \label{ParityPorjectedMBWFcomplex}
\end{align}
where $\alpha_{\pm}$ is the normalization coefficient in association with the positive ($+$) or negative ($-$) parity.

\subsection{Angular-momentum projection}
\label{SecCompProjectionAngMom}
The angular-momentum-projected state $\Ket{\Psi_{\nu MK}^{J}(\mathcal{Z})}$
is defined with the angular-momentum-projection operator $\hat{\mathcal{P}}_{MK}^J$ by
\begin{align}
 \Ket{\Psi_{\nu MK}^{J}(\mathcal{Z})}
 &=
 \hat{\mathcal{P}}_{MK}^J
 \Ket{\Psi_{\nu}(\mathcal{Z})},
 \label{JKMstate}\\
 \hat{\mathcal{P}}_{MK}^J
 &=
 \frac{2J+1}{8\pi^2}
 \int d\Omega D_{MK}^{J*}(\Omega)\hat R(\Omega).
 \label{AMprojector}
\end{align}
Here $M$ is the $z$-component of the total angular momentum $J$ in the laboratory frame, 
where the rotational symmetry is restored,
while $K$ is that for the body-fixed (intrinsic) frame.
The three-dimensional Euler angle $\Omega$ appears as an argument of 
the Wigner $D$-function $D_{MK}^{J}$ and the rotation operator $\hat R$,
which is associated with the rotation in spatial and spin spaces.
The integration over $\Omega$ in Eq.~\eqref{AMprojector} can be performed numerically.
Also we simply denote $\mathcal{Z}$ without the subscript $n$.

The operation of $\hat R$ results in the rotation of the single-particle wave function.
As a result, for the spatial part, we just have to rotate $\vect{Z}_{ij}$ and $\vect{z}_{ij}$ as
\begin{align}
 \vect{Z}_{ij}
 &\to
 \vect{Z}_{ij}^{(R)}
 =
 \frac{1}{2}\left[ \hat{R}(\Omega)\vect{Z}_i +\hat{R}(\Omega)\vect{Z}_j \right],
 \label{rCMijRotcomplex}\\
 \vect{z}_{ij}
 &\to
 \vect{z}_{ij}^{(R)}
 =
 \hat{R}(\Omega)\vect{Z}_i -\hat{R}(\Omega)\vect{Z}_j.
 \label{rrelijRotcomplex}
\end{align}
As regards the rotation of the spin wave function, the coefficients $\alpha_i$ and $\beta_i$ in Eq.~\eqref{spinisospinwf}
are replaced by
\begin{align}
 \begin{pmatrix}
  \alpha_i\\[10pt]
  \beta_i
 \end{pmatrix}
 \to
 \begin{pmatrix}
  \alpha_i'\\[10pt]
  \beta_i'
 \end{pmatrix}
 =
 \begin{pmatrix}
  D_{\frac{1}{2}\frac{1}{2}}^{\frac{1}{2}}(\Omega) & D_{\frac{1}{2},-\frac{1}{2}}^{\frac{1}{2}}(\Omega)\\[10pt]
  D_{-\frac{1}{2},\frac{1}{2}}^{\frac{1}{2}}(\Omega) & D_{-\frac{1}{2},-\frac{1}{2}}^{\frac{1}{2}}(\Omega)
 \end{pmatrix}
 \begin{pmatrix}
  \alpha_i\\[10pt]
  \beta_i
 \end{pmatrix},
 \label{rotspin}
\end{align}
as well as $\alpha_j$ and $\beta_j$ to be replaced by $\alpha_j'$ and $\beta_j'$, respectively.


\section{Potentials derived from chiral effective field theory}
\label{SecChiralpot}
\subsection{Overview}
\label{SecChiralpotov}
We adopt a high-precision two-body potential based on the chiral EFT at N$^3$LO~\cite{ENTEM200293,PhysRevC.66.014002,PhysRevC.68.041001,MACHLEIDT20111} as $v_{2N}$ in Eq.~\eqref{gene2bMEmom2Complex}.
At this order, the potential can be written order by order as
\begin{align}
 v_{2N}\!\left(\vect{p},\vect{p}'\right)
 &=
 v_{2N}^{(\mathrm{LO})}\!\left(\vect{p},\vect{p}'\right)
 +v_{2N}^{(\mathrm{NLO})}\!\left(\vect{p},\vect{p}'\right)
 +v_{2N}^{(\mathrm{N^2LO})}\!\left(\vect{p},\vect{p}'\right)
 +v_{2N}^{(\mathrm{N^3LO})}\!\left(\vect{p},\vect{p}'\right).
 \label{chiralVfull}
\end{align}
The superscript stands for the chiral-expansion power, i.e., $n_\chi=0,2,3$, and $4$ for leading order (LO), 
NLO, next-to-next-to-leading order (N$^2$LO), and N$^3$LO, respectively. 

In this work, we employ the nonlocal regularization with the regulator
\begin{align}
 u_n\!\left(p,p',\Lambda\right)
 &=
 \exp\!\left[-\!\left(\frac{p}{\Lambda}\right)^{\!2n}-\!\left(\frac{p'}{\Lambda}\right)^{\!2n}\right].
 \label{nonlocalreg}
\end{align}
Here $\Lambda$ is the cutoff momentum.
Thus, the potential depends on the relative momenta, $\vect{p}$ and $\vect{p}'$, of the initial and final channels, respectively,
which are related to the average momentum $\vect{Q}$ and transferred momentum $\vect{q}$ by
\begin{align}
  \begin{pmatrix}
  \vect{Q} \\
  \vect{q}
 \end{pmatrix}
 =
 \mathcal{U}
 \begin{pmatrix}
  \vect{p}' \\
  \vect{p}
 \end{pmatrix},
 \label{UtransppqQ}
\end{align}
where $\mathcal{U}$ is given by Eq.~\eqref{Umat1}.
Note that every potential $v_{2N}$ appearing in this paper involves the prefactor $1/(2\pi)^3$, 
which originates from the normalization convention, ${\Braket{\vect{p}|\vect{p}'}=\delta(\vect{p}-\vect{p}')}$.
A similar prefactor for a potential of the chiral three-nucleon force can be found in Refs.~\cite{NavratilFS2007,PhysRevC.98.044305}.

At each order, the potentials consist of the 1$\pi$ exchange term $v_{1\pi}$, 
the two-pion (2$\pi$) exchange term $v_{2\pi}^{(n_\chi)}$, and the contact term $v_{\mathrm{ct}}^{(n_\chi)}$.
We see these contributions in the next sections.

\subsection{Leading order}
\label{SecChiralpotLO}
The LO potential reads
\begin{align}
 v_{2N}^{(\mathrm{LO})}\!\left(\vect{p},\vect{p}'\right)
 &=
 v_{1\pi}\!\left(\vect{p},\vect{p}'\right)+v_{\mathrm{ct}}^{(0)}\!\left(\vect{p},\vect{p}'\right),
 \label{VLO}\\
 v_{1\pi}\!\left(\vect{p},\vect{p}'\right)
 &=
 -\frac{1}{(2\pi)^3}\!\left(\frac{g_A}{2f_\pi}\right)^{\!2}
 u_n\!\left(p,p',\Lambda\right)
 \frac{\left(\vect{\sigma}_1\cdot\vect{q}\right)\left(\vect{\sigma}_2\cdot\vect{q}\right)}{q^2+m_\pi^2}
 \vect{\tau}_1 \cdot \vect{\tau}_2,
 \label{potLO1pi}\\
 v_{\mathrm{ct}}^{(0)}\!\left(\vect{p},\vect{p}'\right)
 &=
 \frac{1}{(2\pi)^3} u_n\!\left(p,p',\Lambda\right)
 \left(C_S+C_T\,\vect{\sigma}_1 \cdot \vect{\sigma}_2\right),
 \label{potCS}
\end{align}
where $g_A$ is the axial vector coupling constant, $f_\pi$ is the pion-decay constant, and $m_\pi$ is the average pion mass, 
as well as the LECs, $C_S$ and $C_T$ at LO.
The spin and isospin operators, $\vect{\sigma}_i$ and $\vect{\tau}_i$ respectively, are represented by the Pauli matrices.

\subsection{Next-to-leading order}
\label{SecChiralpotNLO}
At NLO, the potential is given by
\begin{align}
 v_{2N}^{(\mathrm{NLO})}\!\left(\vect{p},\vect{p}'\right)
 &=
 v_{2\pi}^{(2)}\!\left(\vect{p},\vect{p}'\right)+v_{\mathrm{ct}}^{(2)}\!\left(\vect{p},\vect{p}'\right),
 \label{VNLO}\\
 v_{2\pi}^{(2)}\!\left(\vect{p},\vect{p}'\right)
 &=
 u_n\!\left(p,p',\Lambda\right)
 \Big[
 W_C^{(2)}(q,Q) \, \vect{\tau}_1 \cdot \vect{\tau}_2
 +V_S^{(2)}(q,Q) \, \vect{\sigma}_1 \cdot \vect{\sigma}_2
 \nonumber\\
 &+V_T^{(2)}(q,Q) \left(\vect{\sigma}_1\cdot\vect{q}\right)\left(\vect{\sigma}_2\cdot\vect{q}\right)
 \Big],
 \label{pot2piNLO}\\
 v_{\mathrm{ct}}^{(2)}\!\left(\vect{p},\vect{p}'\right)
 &=
 \frac{1}{(2\pi)^3}u_n\!\left(p,p',\Lambda\right)
 \Big[C_1 q^2 +C_2 Q^2 +\left(C_3 q^2 +C_4 Q^2\right)\vect{\sigma}_1 \cdot \vect{\sigma}_2
 \nonumber\\
 &+C_5\left[-i\vect{S}\cdot\left(\vect{q}\times\vect{Q}\right)\right]
  +C_6\left(\vect{\sigma}_1\cdot\vect{q}\right)\left(\vect{\sigma}_2\cdot\vect{q}\right)
  +C_7\left(\vect{\sigma}_1\cdot\vect{Q}\right)\left(\vect{\sigma}_2\cdot\vect{Q}\right)\Big],
 \label{potNLOct}
\end{align}
where $\vect{S}=(\vect{\sigma}_1+\vect{\sigma}_2)/2$ and $C_i$ are the LECs at NLO.
The $2\pi$ term contains
\begin{align}
 W_C^{(2)}(q,Q)
 &=
 -\frac{L(q)}{3(4\pi)^5 f_\pi^4}
 \nonumber\\
 &\times
 \left[4m_\pi^2\left(5g_A^4-4g_A^2-1\right)
 +q^2\left(23g_A^4-10g_A^2-1\right)
 +48g_A^4 m_\pi^4w^{-2}\right],
 \label{WCNLO}\\
 V_S^{(2)}(q,Q)
 &=
 \frac{6g_A^4}{(4\pi)^5 f_\pi^4}q^2L(q),
 \label{VSNLO}\\
 V_T^{(2)}(q,Q)
 &=
 -\frac{6g_A^4}{(4\pi)^5 f_\pi^4}L(q),
 \label{VTNLO}
\end{align}
with
\begin{align}
 w
 &=
 \sqrt{4m_\pi^2 +q^2},
 \label{2piw}\\
 L(q)
 &=
 \frac{w}{q}\ln\frac{w+q}{2m_\pi}.
 \label{2piLq}
\end{align}

\subsection{Next-to-next-to-leading order}
\label{SecChiralpotN2LO}
At N$^2$LO, we have the potential
\begin{align}
 v_{2N}^{(\mathrm{N^2LO})}\!\left(\vect{p},\vect{p}'\right)
 &=
 v_{2\pi}^{(3)}\!\left(\vect{p},\vect{p}'\right)
 \nonumber\\
 &=
 u_n\!\left(p,p',\Lambda\right)
 \Bigg[V_C^{(3)}(q,Q)+W_C^{(3)}(q,Q) \, \vect{\tau}_1 \cdot \vect{\tau}_2
 \nonumber\\
 & +\left[V_S^{(3)}(q,Q)+W_S^{(3)}(q,Q) \, \vect{\tau}_1 \cdot \vect{\tau}_2 \right]
 \vect{\sigma}_1 \cdot \vect{\sigma}_2
 \nonumber\\
 &+\left[V_{LS}^{(3)}(q,Q) +W_{LS}^{(3)}(q,Q) \, \vect{\tau}_1 \cdot \vect{\tau}_2\right]
 \left[-i\vect{S}\cdot\left(\vect{q}\times \vect{Q}\right)\right]
 \nonumber\\
 &+\left[V_{T}^{(3)}(q,Q) +W_T^{(3)}(q,Q) \, \vect{\tau}_1 \cdot \vect{\tau}_2\right]
 \left(\vect{\sigma}_1\cdot\vect{q}\right)\left(\vect{\sigma}_2\cdot\vect{q}\right)
 \Bigg].
 \label{pot2piN2LO}
\end{align}
Here, the potentials involve the central terms,
\begin{align}
 V_C^{(3)}(q,Q)
 &=
 \frac{6g_A^2}{(4\pi)^4 f_\pi^4}
 \Bigg[
 \frac{g_A^2 m_\pi^5}{16 m_N w^2}
 -\left\{2m_\pi^2\left(2c_1-c_3\right) -q^2\left(c_3 +\frac{3g_A^2}{16m_N}\right)\right\}
 \tilde w^2 A(q)
 \nonumber\\
 &-\frac{g_A^2}{16 m_N}\left\{m_\pi w^2 +\tilde w^4 A(q)\right\}
 \Bigg],
 \label{potVCN2LO2}\\
 W_C^{(3)}(q,Q)
 &=
 \frac{g_A^2}{4(4\pi)^4 m_N f_\pi^4}
 \Bigg[
  3g_A^2 m_\pi^5 w^{-2}
 -\left\{4m_\pi^2 +2q^2-g_A^2\left(4m_\pi^2+3q^2\right)\right\}
 \tilde w^2 A(q)
 \nonumber\\
 &+g_A^2\left\{m_\pi w^2 +\tilde w^4 A(q)\right\}
 \Bigg],
 \label{potWCN2LO2}\\
 V_S^{(3)}(q,Q)
 &=
 -\frac{3g_A^4}{(8\pi)^4 m_N f_\pi^4}
 \Bigg[
  3q^2 \tilde w^2 A(q)
 +q^2 \left\{m_\pi +w^2 A(q)\right\}
 \Bigg],
 \label{potVSN2LO2}\\
 W_S^{(3)}(q,Q)
 &=
 \frac{g_A^2}{(4\pi)^4 f_\pi^4}
 \Bigg[
  q^2 A(q)
 \left\{\left(c_4+\frac{1}{4m_N}\right)w^2
 -\frac{g_A^2}{8m_N}\left(10m_\pi^2+3q^2\right)\right\}
 \nonumber\\
 &+\frac{g_A^2}{8m_N}q^2 \left\{m_\pi +w^2 A(q)\right\}
 \Bigg],
 \label{potWSN2LO2}
\end{align}
the SO terms,
\begin{align}
 V_{LS}^{(3)}(q,Q)
 &=
 \frac{3g_A^4}{(4\pi)^4 m_N f_\pi^4}
 \tilde w^2 A(q),
 \label{potVLSN2LO2}\\
 W_{LS}^{(3)}(q,Q)
 &=
 \frac{g_A^2\left(1-g_A^2\right)}{(4\pi)^4 m_N f_\pi^4}
 w^2 A(q),
 \label{potWLSN2LO2}
\end{align}
and the tensor terms,
\begin{align}
 V_T^{(3)}(q,Q)
 &=
 -\frac{1}{q^2} V_S^{(3)}(q,Q),
 \label{potVTN2LO2}\\
 W_T^{(3)}(q,Q)
 &=
 -\frac{1}{q^2} W_S^{(3)}(q,Q),
\end{align}
with
\begin{align}
 \tilde w
 &=
 \sqrt{2m_\pi^2 +q^2},
 \label{2piwtil}\\
 A(q)
 &=
 \frac{1}{2q}\arctan\frac{q}{2m_\pi}.
 \label{2piAq}
\end{align}
The LECs $c_i$ enter the $2\pi$ term at N$^2$LO,
and $m_N$ denotes the average nucleon mass.

\subsection{Next-to-next-to-next-to-leading order}
\label{SecChiralpotN3LO}
At N$^3$LO, the potentials are represented as
\begin{align}
 v_{2N}^{(\mathrm{N^3LO})}\!\left(\vect{p},\vect{p}'\right)
 &=
 v_{2\pi}^{(4)}\!\left(\vect{p},\vect{p}'\right)+v_{\mathrm{ct}}^{(4)}\!\left(\vect{p},\vect{p}'\right),
 \label{VN3LO}\\
 v_{2\pi}^{(4)}\!\left(\vect{p},\vect{p}'\right)
 &=
 u_n\!\left(p,p',\Lambda\right)
 \Bigg[
 V_{C}^{(4)}(q,Q)+W_{C}^{(4)}(q,Q) \vect{\tau}_1 \cdot \vect{\tau}_2
 \nonumber\\
 &+V_{S}^{(4)}(q,Q) \vect{\sigma}_1 \cdot \vect{\sigma}_2
 +W_{S}^{(4)}(q,Q) \left(\vect{\sigma}_1 \cdot \vect{\sigma}_2\right) \left(\vect{\tau}_1 \cdot \vect{\tau}_2\right)
 \nonumber\\
 &+\left[V_{LS}^{(4)}(q,Q) +W_{LS}^{(4)}(q,Q) \, \vect{\tau}_1 \cdot \vect{\tau}_2\right]
 \left[-i\vect{S}\cdot\left(\vect{q}\times \vect{Q}\right)\right]
 \nonumber\\
 &+\left[V_T^{(4)}(q,Q)+W_T^{(4)}(q,Q)\vect{\tau}_1 \cdot \vect{\tau}_2\right]
 \left(\vect{\sigma}_1\cdot\vect{q}\right)\left(\vect{\sigma}_2\cdot\vect{q}\right)
 \nonumber\\
 &+V_{\sigma L}^{(4)}(q,Q)
 \left[\vect{\sigma}_1\cdot\left(\vect{q}\times\vect{Q}\right)\right]
 \left[\vect{\sigma}_2\cdot\left(\vect{q}\times\vect{Q}\right)\right]
 \Bigg],
 \label{potN3LO2pi}\\
 v_{\mathrm{ct}}^{(4)}\!\left(\vect{p},\vect{p}'\right)
 &=
 \frac{1}{(2\pi)^3}
 u_n\!\left(p,p',\Lambda\right)
 \Bigg[
 D_1 q^4 +D_2 Q^4 +D_3q^2Q^2 +D_4 \left(\vect{q}\times\vect{Q}\right)^2
 \nonumber\\
 &+
 \left[D_5 q^4 +D_6 Q^4 +D_7q^2Q^2 +D_8 \left(\vect{q}\times\vect{Q}\right)^2\right]
 \vect{\sigma}_1 \cdot \vect{\sigma}_2
 \nonumber\\
 &+
 \left[D_9 q^2 +D_{10} Q^2\right]
 \left[-i\vect{S}\cdot\left(\vect{q}\times\vect{Q}\right)\right]
 \nonumber\\
 &+
 \left[D_{11} q^2 +D_{12} Q^2\right]
 \left(\vect{\sigma}_1\cdot\vect{q}\right)\left(\vect{\sigma}_2\cdot\vect{q}\right)
 \nonumber\\
 &+
 \left[D_{13} q^2 +D_{14} Q^2\right]
 \left(\vect{\sigma}_1\cdot\vect{Q}\right)\left(\vect{\sigma}_2\cdot\vect{Q}\right)
 \nonumber\\
 &+
 D_{15} \left[\vect{\sigma}_1\cdot\left(\vect{q}\times\vect{Q}\right)\right]
 \left[\vect{\sigma}_2\cdot\left(\vect{q}\times\vect{Q}\right)\right]\Bigg].
 \label{potN3LOct}
\end{align}
The LECs of the contact terms at N$^3$LO are $D_i$.

Following Ref.~\cite{MACHLEIDT20111}, the 2$\pi$-exchange potential at N$^3$LO
is categorized into several terms, i.e., $c_i^2$, $c_i/{m_N}$, and
$m_N^{-2}$ contributions in association with one-loop diagrams, and also
two-loop (2L) contributions, where $c_i$ stands for one of the LECs.
Thus, potentials in Eq.~\eqref{potN3LO2pi} can be decomposed further.
The purely central terms of $v_{2\pi}^{(4)}$ read
\begin{align}
 V_C^{(4)}(q,Q)
 &=
 V_C^{(c_i^2)}(q,Q)
 +V_C^{(c_i/m_N)}(q,Q)
 +V_C^{(m_N^{-2})}(q,Q)
 +V_C^{(\mathrm{2L})}(q,Q),
 \label{2piN3LOVC}\\
 V_C^{(c_i^2)}(q,Q)
 &=
 \frac{3}{4(2\pi)^5 f_\pi^4}
 L(q)
 \left[\left(\frac{c_2}{6}w^2+c_3\tilde w^2 -4c_1m_\pi^2\right)^2
 +\frac{c_2^2}{45}w^4\right],
 \label{potVCN3LO2ci2}\\
 V_C^{(c_i/m_N)}(q,Q)
 &=
 -\frac{g_A^2}{8(2\pi)^5 m_N f_\pi^4}
  L(q)
 \Bigg[
 \left\{(c_2-6c_3)q^4
 +4(6c_1+c_2-3c_3)q^2m_\pi^2
 \right.
 \nonumber\\
 &\left.
 +6(c_2-2c_3)m_\pi^4
 +24(2c_1+c_3)m_\pi^6w^{-2}\right\}
 \Bigg],
 \label{potVCN3LO2cimN}\\
 V_C^{(m_N^{-2})}(q,Q)
 &=
 -\frac{g_A^4}{8(2\pi)^5 m_N^2 f_\pi^4}
 \Bigg[
 L(q)
 \left(2m_\pi^8 w^{-4} +8m_\pi^6w^{-2}-q^4-2m_\pi^4\right)
 +\frac{1}{2}m_\pi^6w^{-2}
 \Bigg],
 \label{potVCN3LO2mN-2}\\
 V_C^{(\mathrm{2L})}(q,Q)
 &=
 \frac{3g_A^4}{8(4\pi)^5 f_\pi^6}
 \tilde w^2 A(q)
 \left[\left(m_\pi^2+2q^2\right)\left(2m_\pi+\tilde w^2 A(q)\right)
 +4g_A^2 m_\pi \tilde w^2\right],
 \label{potVCN3LO22l}
 \end{align}
as well as those with the isospin dependent terms,
\begin{align}
 W_C^{(4)}(q,Q)
 &=
 W_C^{(c_i/m_N)}(q,Q)
 +W_C^{(m_N^{-2})}(q,Q)
 +W_C^{(\mathrm{2L})}(q,Q),
 \label{2piN3LOWC}\\
 W_C^{(c_i/m_N)}(q,Q)
 &=
 -\frac{1}{48(2\pi)^5 m_N f_\pi^4}
 c_4 q^2 L(q)
 \left[g_A^2\left(8m_\pi^2+5q^2\right)+w^2\right],
 \label{potWCN3LO2cimN}\\
 W_C^{(m_N^{-2})}(q,Q)
 &=
 -\frac{1}{6(4\pi)^5 m_N^2 f_\pi^4}
 L(q)
 \nonumber\\
 &\times \Bigg[
 \Big[
 8g_A^2\left\{\frac{3}{2}q^4+3m_\pi^2q^2+3m_\pi^4-6m_\pi^6w^{-2}
 -Q^2\left(8m_\pi^2+5q^2\right)\right\}
 \nonumber\\
 &+4g_A^4\left\{
 Q^2\left(20m_\pi^2+7q^2-16m_\pi^4w^{-2}\right)
 \right.
 \nonumber\\
 & \left.
 +16m_\pi^8w^{-4}+12m_\pi^6w^{-2}
 -4m_\pi^4q^2w^{-2}-5q^4-6m_\pi^2q^2-6m_\pi^4
 \right\}
 -4Q^2w^2
 \Big]
 \nonumber\\
 &+16g_A^4m_\pi^6w^{-2}
 \Bigg],
 \label{potWCN3LO2mN-2}\\
 W_C^{(\mathrm{2L})}(q,Q)
 &=
 \frac{1}{9(4\pi)^7 f_\pi^6}
 L(q)
 \nonumber\\
 &\times
 \Bigg[
 192\pi^2f_\pi^2w^2\bar d_3
 \left\{2g_A^2\tilde w^2-\frac{3}{5}\left(g_A^2-1\right) w^2\right\}
 +\left\{6g_A^2\tilde w^2-\left(g_A^2-1\right)w^2\right\}
 \nonumber\\
 &\times \Big[
 384\pi^2f_\pi^2
 \left\{\tilde w^2 \left(\bar d_1 +\bar d_2\right)+4m_\pi^2 \bar d_5 \right\}
 +L(q)\left\{4m_\pi^2\left(1+2g_A^2\right)+q^2\left(1+5g_A^2\right)\right\}
 \nonumber\\
 & \left.
 -\frac{1}{3}q^2\left(5+13g_A^2\right)
 -8m_\pi^2\left(1+2g_A^2\right)
 \right]\Bigg].
 \label{potWCN3LO22l}
\end{align}
Here, $\bar d_i$ are the LECs.
The spin-spin terms are given by
\begin{align}
 V_S^{(4)}(q,Q)
 &=
 V_S^{(m_N^{-2})}(q,Q)
 +V_S^{(\mathrm{2L})}(q,Q),
 \label{2piN3LOVS}\\
 V_S^{(m_N^{-2})}(q,Q)
 &=
 -\frac{g_A^4}{8(2\pi)^5 m_N^2 f_\pi^4}
 q^2 L(q)
 \left[Q^2+\frac{5}{8}q^2+m_\pi^4w^{-2}\right],
 \label{potVSN3LO2mN-2}\\
 V_S^{(\mathrm{2L})}(q,Q)
 &=
 \frac{g_A^2}{8(2\pi)^5 f_\pi^4}
 \left(\bar d_{14}-\bar d_{15}\right)
 q^2 w^2 L(q),
 \label{potVSN3LO22l}
\end{align}
and
\begin{align}
 W_S^{(4)}(q,Q)
 &=
 V_S^{(c_i^2)}(q,Q)
 +V_S^{(c_i/m_N)}(q,Q)
 +V_S^{(m_N^{-2})}(q,Q)
 +V_S^{(\mathrm{2L})}(q,Q),
 \label{2piN3LOWS}\\
 W_S^{(c_i^2)}(q,Q)
 &=
 -\frac{4}{3(4\pi)^5 f_\pi^4}
 c_4^2 q^2 w^2 L(q),
 \label{potWSN3LO2ci2}\\
 W_S^{(c_i/m_N)}(q,Q)
 &=
 \frac{1}{48(2\pi)^5 m_N f_\pi^4}
 c_4 q^2 L(q)
 \left[g_A^2\left(16m_\pi^2+7q^2\right)-w^2\right],
 \label{potWSN3LO2cimN}\\
 W_S^{(m_N^{-2})}(q,Q)
 &=
 -\frac{1}{12(4\pi)^5 m_N^2 f_\pi^4}
 q^2 L(q)
 \nonumber\\
 &\times
 \left[
 4g_A^4\left(7m_\pi^2+\frac{17}{4}q^2+4m_\pi^4w^{-2}\right)
 -32g_A^2\left(m_\pi^2+\frac{7}{16}q^2\right)+w^2
 \right],
 \label{potWSN3LO2mN-2}\\
 W_S^{(\mathrm{2L})}(q,Q)
 &=
 -\frac{g_A^4}{16(4\pi)^5 f_\pi^6}
 q^2 w^2 A(q)
 \left[ w^2 A(q) +2m_\pi\left(1+2g_A^2\right) \right].
 \label{FWSN3LO2l}
\end{align}
The SO terms read
\begin{align}
 V_{LS}^{(4)}(q,Q)
 &=
 V_{LS}^{(c_i/m_N)}(q,Q)
 +V_{LS}^{(m_N^{-2})}(q,Q),
 \label{2piN3LOVLS}\\
 V_{LS}^{(c_i/m_N)}(q,Q)
 &=
 \frac{g_A^2}{2(2\pi)^5 m_N f_\pi^4}
 c_2 w^2 L(q),
 \label{potVLSN3LOcimN2}\\
 V_{LS}^{(m_N^{-2})}(q,Q)
 &=
 \frac{g_A^4}{(2\pi)^5 m_N^2 f_\pi^4}
 L(q)
 \left[\frac{11}{32}q^2+m_\pi^4w^{-2}\right],
 \label{potVLSN3LOmN-22}
\end{align}
as well as
\begin{align}
 W_{LS}^{(4)}(q,Q)
 &=
 W_{LS}^{(c_i/m_N)}(q,Q)
 +W_{LS}^{(m_N^{-2})}(q,Q),
 \label{2piN3LOWLS}\\
 W_{LS}^{(c_i/m_N)}(q,Q)
 &=
 -\frac{1}{12(2\pi)^5 m_N f_\pi^4}
 c_4 L(q)
 \left[g_A^2\left(8m_\pi^2+5q^2\right)+w^2\right],
 \label{potWLSN3LOcimN2}\\
 W_{LS}^{(m_N^{-2})}(q,Q)
 &=
 \frac{1}{2(4\pi)^5 m_N^2 f_\pi^4}
 L(q)
 \nonumber\\
 &\times
 \left[
 16g_A^2\left(m_\pi^2+\frac{3}{8}q^2\right)
 +\frac{4}{3}g_A^4\left(4m_\pi^4w^{-2}-\frac{11}{4}q^2-9m_\pi^2\right)
 -w^2\right].
 \label{potWLSN3LOmN-22}
\end{align}
The tensor terms are expressed by
\begin{align}
 V_T^{(4)}(q,Q)
 &=
 V_T^{(m_N^{-2})}(q,Q)
 +V_T^{(\mathrm{2L})}(q,Q),
 \label{2piN3LOVT}\\
 V_T^{(X)}(q,Q)
 &=
 -\frac{1}{q^2} V_S^{(X)}(q,Q),
 \label{potVTN3LO2general}
\end{align}
and
\begin{align}
 W_T^{(4)}(q,Q)
 &=
 W_T^{(c_i^2)}(q,Q)
 +W_T^{(c_i/m_N)}(q,Q)
 +W_T^{(m_N^{-2})}(q,Q)
 +W_T^{(\mathrm{2L})}(q,Q),
 \label{2piN3LOWT}\\
 W_T^{(X)}(q,Q)
 &=
 -\frac{1}{q^2} W_S^{(X)}(q,Q).
 \label{potWTN3LO2general}
\end{align}
Here $X$ is a representative of $c_i^2$, $c_i/m_N$, $m_N^{-2}$, and 2L.
The $V_{\sigma L}^{(4)}$ term in Eq.~\eqref{potN3LO2pi} also behaves as a tensor force,
and the potential is given by
\begin{align}
 V_{\sigma L}^{(4)}(q,Q)
 &=
 V_{\sigma L}^{(m_N^{-2})}(q,Q)
 =
 \frac{g_A^4}{8(2\pi)^5 m_N^2 f_\pi^4}
 L(q).
 \label{2piN3LOVsigmaL}
\end{align}


\section{Multipole-expansion function}
\label{SecMPE}
The purpose of the MPE is to express the chiral potential, which is originally given as a function of $\vect{q}$ and $\vect{Q}$,
in terms of $\vect{p}$ and $\vect{p}'$.
This is relevant to the nonlocal regularization.

In this section, we derive the explicit form of the MPE functions,
$f_{L}^{(C)}$, $f_{LL'L_1L_2K}^{(LS)}$, $f_{\lambda_0 K}^{(T)}$, and $f_{L_qL_QK}^{(\sigma L)}$.
In general, the MPE function, which depends on $p$ and $p'$, is given by integration over $x$,
where $x$ is defined by
\begin{align}
 x=\frac{\vect{p}\cdot\vect{p}'}{pp'}.
 \label{anglexpp}
\end{align}
It appears in $q$ and $Q$ 
as $q=\sqrt{p^2+p'^2-2pp'x}$ and $Q=\sqrt{p^2+p'^2+2pp'x}/2$, respectively.
This integration needs to be performed numerically for the $1\pi$ and $2\pi$ terms,
while it can be calculated analytically for the contact terms.

\subsection{Central contributions}
\label{SecMPEcentral}
The central contributions are in association with the operators, 
$\mathbb{1}$, $\vect{\sigma}_1\cdot\vect{\sigma}_2$, $\vect{\tau}_1\cdot\vect{\tau}_2$, and 
$\left(\vect{\sigma}_1\cdot\vect{\sigma}_2\right)\left(\vect{\tau}_2\cdot\vect{\tau}_2\right)$.
The corresponding terms of the chiral interaction can be found in Table~\ref{tableChiralSpinIso}.
It is convenient to separate $f_{L}^{(C)}$ of the contact terms from that of the $2\pi$ terms.
Furthermore, we put an additional symbol explicitly in the superscript of the MPE function
to distinguish each term.
Thus, we find $f_{L}^{(C)}$ of the contact terms as
\begin{align}
 f_{L}^{(C;C_S)}\!\left(p,p'\right)
 &=
 \frac{C_S}{C_T}
 f_{L}^{(C;C_T)}\!\left(p,p'\right)
 =
 \frac{2}{\pi}C_S \delta_{L0},
 \label{MPECSCT}\\
 f_{L}^{(C;C_1)}\!\left(p,p'\right)
 &=
 \frac{4C_1}{C_2}(-)^L
 f_{L}^{(C;C_2)}\!\left(p,p'\right)
 =
 \frac{C_1}{C_3}
 f_{L}^{(C;C_3)}\!\left(p,p'\right)
 =
 \frac{4C_1}{C_4}(-)^L
 f_{L}^{(C;C_4)}\!\left(p,p'\right)
 \nonumber\\
 &=
 \frac{2}{\pi} C_1\left[\left(p^2+p'^2\right)\delta_{L 0}-2pp'\delta_{L 1}\right],
 \label{MPEC1C2C3C4}\\
 f_{L}^{(C;D_1)}\!\left(p,p'\right)
 &=
 \frac{16D_1}{D_2}(-)^L
 f_{L}^{(C;D_2)}\!\left(p,p'\right)
 =
 \frac{D_1}{D_5}
 f_{L}^{(C;D_5)}\!\left(p,p'\right)
 =
 \frac{16D_1}{D_6}(-)^L
 f_{L}^{(C;D_6)}\!\left(p,p'\right)
 \nonumber\\
 &=
 \frac{2}{\pi}D_1\left[\left\{\left(p^2+p'^2\right)^2+\frac{4}{3}p^2p'^2\right\}\delta_{L 0}
 -4 \left(p^2+p'^2\right)pp'\delta_{L 1}
 +\frac{8}{3} p^2p'^2\delta_{L 2}\right],
 \label{MPED1D2D5D6}\\
 f_{L}^{(C;D_3)}\!\left(p,p'\right)
 &=
 \frac{D_3}{D_7}
 f_{L}^{(C;D_7)}\!\left(p,p'\right)
 \nonumber\\
 &=
 \frac{1}{2\pi}D_3
 \left[\left\{\left(p^2+p'^2\right)^2
 -\frac{4}{3}p^2p'^2\right\}\delta_{L 0}
 -\frac{8}{3}p^2p'^2\delta_{L 2} \right],
 \label{MPED3}\\
 f_{L}^{(C;D_4)}\!\left(p,p'\right)
 &=
 \frac{2D_4}{3D_3} f_{L}^{(C;D_3)}\!\left(p,p'\right)
 +D_4 \mathcal{G}_{L}\!\left(p,p'\right),
 \label{MPED4}\\
 f_{L}^{(C;D_8)}\!\left(p,p'\right)
 &=
 \frac{2D_8}{3D_7} f_{L}^{(C;D_7)}\!\left(p,p'\right)
 +D_8 \mathcal{G}_{L}\!\left(p,p'\right),
 \label{MPED8}\\
 \mathcal{G}_{L}\!\left(p,p'\right)
 &=
 -\frac{1}{15\pi}
 \sum_{\Lambda_q=0}^2
 \sum_{\Lambda_Q=0}^2
 (-)^{\Lambda_Q}
 \widehat{2\!-\!\Lambda_q}\widehat{2\!-\!\Lambda_Q}
 \left[\binom{5}{2\Lambda_q}\binom{5}{2\Lambda_Q}\right]^{\!\frac{1}{2}}
 \nonumber\\
 &\times
 \left( \Lambda_q 0 \Lambda_Q 0 | L 0 \right)
 \left( 2\!-\!\Lambda_q, 0, 2\!-\!\Lambda_Q, 0 | L 0 \right)
 \begin{Bmatrix}
  \Lambda_q   & 2\!-\!\Lambda_q & 2 \\
  2\!-\!\Lambda_Q & \Lambda_Q   & L
 \end{Bmatrix}
 \nonumber\\
 &\times
 p^{\Lambda_q+\Lambda_Q} p'^{4-\Lambda_q-\Lambda_Q}.
 \label{MPEhpp}
\end{align}

As regards the pion-exchange terms, $f_{L}^{(C)}$ can be always written as
\begin{align}
 f_{L}^{(C)}\!\left(p,p'\right)
 &=
 \frac{(4\pi)^2}{2}\hat L^2
 \int_{-1}^1 dx P_L(x)
 G^{(C)}(q,Q),
 \label{MPEfunpigeneralC}
\end{align}
where $G^{(C)}(q,Q)$ is a representative for each term defined in Appendix~\ref{SecChiralpot}:
\begin{align}
 G^{(C)}(q,Q)
 =
 \left\{
 \begin{aligned}
  &W_C^{(2)}(q,Q) &&\qquad (\text{$\tau$ exchange at NLO}),\\
  &V_S^{(2)}(q,Q) &&\qquad (\text{$\sigma$ exchange at NLO}),\\
  &V_C^{(3)}(q,Q) &&\qquad (\text{Purely central at N$^2$LO}),\\
  &V_S^{(3)}(q,Q) &&\qquad (\text{$\sigma$ exchange at N$^2$LO}),\\
  &W_C^{(3)}(q,Q) &&\qquad (\text{$\tau$ exchange at N$^2$LO}),\\
  &W_S^{(3)}(q,Q) &&\qquad (\text{$\sigma$/$\tau$ exchange at N$^2$LO}),\\
  &V_C^{(c_i^2)}(q,Q) &&\qquad (\text{Purely central for N$^3$LO-$c_i^2$ term}),\\
  &V_C^{(c_i/m_N)}(q,Q) &&\qquad (\text{Purely central for N$^3$LO-$c_i/m_N$ term}),\\
  &V_C^{(m_N^{-2})}(q,Q) &&\qquad (\text{Purely central for N$^3$LO-$m_N^{-2}$ term}),\\
  &V_C^{(\mathrm{2L})}(q,Q) &&\qquad (\text{Purely central for N$^3$LO-2L term}),\\
  &W_C^{(c_i/m_N)}(q,Q) &&\qquad (\text{$\tau$ exchange for N$^3$LO-$c_i/m_N$ term}),\\
  &W_C^{(m_N^{-2})}(q,Q) &&\qquad (\text{$\tau$ exchange for N$^3$LO-$m_N^{-2}$ term}),\\
  &W_C^{(\mathrm{2L})}(q,Q) &&\qquad (\text{$\tau$ exchange for N$^3$LO-2L term}),\\
  &V_S^{(m_N^{-2})}(q,Q) &&\qquad (\text{$\sigma$ exchange for N$^3$LO-$m_N^{-2}$ term}),\\
  &V_S^{(\mathrm{2L})}(q,Q) &&\qquad (\text{$\sigma$ exchange for N$^3$LO-2L term}),\\
  &W_S^{(c_i^2)}(q,Q) &&\qquad (\text{$\sigma$/$\tau$ exchange for N$^3$LO-$c_i^2$ term}),\\
  &W_S^{(c_i/m_N)}(q,Q) &&\qquad (\text{$\sigma$/$\tau$ exchange for N$^3$LO-$c_i/m_N$ term}),\\
  &W_S^{(m_N^{-2})}(q,Q) &&\qquad (\text{$\sigma$/$\tau$ exchange for N$^3$LO-$m_N^{-2}$ term}),\\
  &W_S^{(\mathrm{2L})}(q,Q) &&\qquad (\text{$\sigma$/$\tau$ exchange for N$^3$LO-2L term}).
 \end{aligned}
 \right.
 \label{GCqQ}
\end{align}

\subsection{Spin-orbit contributions}
\label{SecMPESO}
Here again, the contact and $2\pi$ terms of the SO contributions are separately formulated.
The MPE function of the SO contributions of the contact terms reads
\begin{align}
 f_{LL'L_1L_2K}^{(LS;C_5)}\!\left(p,p'\right)
 &=
 4\sqrt{3} C_5 
 \begin{Bmatrix}
  \lambda_q & 1\!-\!\lambda_q & 1 \\
  \lambda_Q & 1\!-\!\lambda_Q & 1 \\
  L         & L'              & 1
 \end{Bmatrix}
 \delta_{LL_1} \delta_{L'L_2} \delta_{K1},
 \label{MPEC5}\\
 f_{LL'L_1L_2K}^{(LS;D_9)}\!\left(p,p'\right)
 &=
 \frac{4D_9}{D_{10}}(-)^K
 f_{LL'L_1L_2K}^{(LS;D_{10})}\!\left(p,p'\right)
 \nonumber\\
 &=
 -4\sqrt{3} D_9
 \hat{L}_1\hat{L}_2
 \left( L_1 0 K 0 | L 0 \right)
 \left( L_2 0 K 0 | L' 0 \right)
 \nonumber\\
 &\times
 \begin{Bmatrix}
  L   & L'  & 1 \\
  L_2 & L_1 & K
 \end{Bmatrix}
 \begin{Bmatrix}
  \lambda_q & 1\!-\!\lambda_q & 1 \\
  \lambda_Q & 1\!-\!\lambda_Q & 1 \\
  L_1       & L_2             & 1
 \end{Bmatrix}
 \left[ \left(p^2+p'^2\right)\delta_{K 0} -2pp'\delta_{K 1}\right],
 \label{MPED9D10}
\end{align}
where we put the corresponding LECs in the superscripts.

The $2\pi$-SO contributions are characterized by
\begin{align}
 f_{LL'L_1L_2K}^{(LS)}\!\left(p,p'\right)
 &=
 -\frac{\sqrt{3}(4\pi)^3}{4}
 \hat L^2
 \left( L_1 0 K 0 | L 0 \right)
 \left( L_2 0 K 0 | L' 0 \right)
 \nonumber\\
 &\times
 \begin{Bmatrix}
  L   & L'  & 1 \\
  L_2 & L_1 & K
 \end{Bmatrix}
 \begin{Bmatrix}
  \lambda_q & 1\!-\!\lambda_q & 1 \\
  \lambda_Q & 1\!-\!\lambda_Q & 1 \\
  L_1       & L_2             & K
 \end{Bmatrix}
 \nonumber\\
 &\times
 \int_{-1}^1 dx P_K(x)
 G^{(LS)}(q,Q),
 \label{MPEfunpigeneralLS}
\end{align}
and $G^{(LS)}$ is given by
\begin{align}
 G^{(LS)}(q,Q)
 =
 \left\{
 \begin{aligned}
  &V_{LS}^{(3)}(q,Q) &&\qquad (\text{SO at N$^2$LO}),\\
  &W_{LS}^{(3)}(q,Q) &&\qquad (\text{SO with $\tau$ exchange at N$^2$LO}),\\
  &V_{LS}^{(c_i/m_N)}(q,Q) &&\qquad (\text{SO for N$^3$LO-$c_i/m_N$ term}),\\
  &V_{LS}^{(m_N^{-2})}(q,Q) &&\qquad (\text{SO for N$^3$LO-$m_N^{-2}$ term}),\\
  &W_{LS}^{(c_i/m_N)}(q,Q) &&\qquad (\text{SO with $\tau$ exchange for N$^3$LO-$c_i/m_N$ term}),\\
  &W_{LS}^{(m_N^{-2})}(q,Q) &&\qquad (\text{SO with $\tau$ exchange for N$^3$LO-$m_N^{-2}$ term}).
 \end{aligned}
 \right.
 \label{GLSqQ}
\end{align}
See Appendix~\ref{SecChiralpot} for explicit form of $G^{(LS)}$.

\subsection{Tensor contributions}
\label{SecMPEtensor}
By explicitly putting LECs in the superscript of $f_{\lambda_0K}^{(T)}$
to distinguish each term of the chiral interaction, 
we can write the MPE function for the tensor contributions of the contact terms:
\begin{align}
 f_{\lambda_0K}^{(T;C_6)}\!\left(p,p'\right)
 &=
 \frac{4C_6}{C_7}
 (-)^K
 f_{\lambda_0K}^{(T;C_7)}\!\left(p,p'\right)
 \nonumber\\
 &=
 24C_6
 \left[
 \left\{
 \left(p^2+p'^2\right)\delta_{K 0}-2pp'\delta_{K 1}
 \right\}\delta_{\lambda_0 0}
 +
 \delta_{\lambda_0 2}\delta_{K 0}
 \right],
 \label{MPEC6C7}\\
 f_{\lambda_0K}^{(T;D_{11})}\!\left(p,p'\right)
 &=
 24D_{11}
 \Bigg[
 \left[\left\{\left(p^2+p'^2\right)^2 +\frac{4}{3}p^2p'^2\right\}\delta_{K0}
 -4\left(p^2+p'^2\right)pp'\delta_{K1}
 +\frac{8}{3}p^2p'^2\delta_{K2}
 \right]\delta_{\lambda_0 0}
 \nonumber\\
 &+\left[
 \left(p^2+p'^2\right)\delta_{K0}-2pp'\delta_{K1}
 \right]\delta_{\lambda_0 2}
 \Bigg],
 \label{MPED11}\\
 f_{\lambda_0K}^{(T;D_{12})}\!\left(p,p'\right)
 &=
 6D_{12}
 \Bigg[
 \left[\left\{\left(p^2+p'^2\right)^2 -\frac{4}{3}p^2p'^2\right\}\delta_{K0}
 -\frac{8}{3}p^2p'^2\delta_{K2}
 \right]\delta_{\lambda_0 0}
 \nonumber\\
 &+\left[
 \left(p^2+p'^2\right)\delta_{K0}+2pp'\delta_{K1}
 \right]\delta_{\lambda_0 2}
 \Bigg],
 \label{MPED12}\\
 f_{\lambda_0K}^{(T;D_{13})}\!\left(p,p'\right)
 &=
 D_{13}
 \left[
 \frac{4}{D_{12}}f_{\lambda_0K}^{(T;D_{12})}\!\left(p,p'\right)\delta_{\lambda_0 0}
 +
 \frac{1}{D_{11}}f_{\lambda_0K}^{(T;D_{11})}\!\left(p,p'\right)\delta_{\lambda_0 2}
 \right],
 \label{MPED13}\\
 f_{\lambda_0K}^{(T;D_{14})}\!\left(p,p'\right)
 &=
 D_{14}
 \left[
 \frac{1}{16D_{11}}f_{\lambda_0K}^{(T;D_{11})}\!\left(p,p'\right)\delta_{\lambda_0 2}
 +
 \frac{4}{D_{12}}f_{\lambda_0K}^{(T;D_{12})}\!\left(p,p'\right)\delta_{\lambda_0 0}
 \right].
 \label{MPED14}
\end{align}

A form similar to Eq.~\eqref{MPEfunpigeneralC} can be found for $f_{\lambda_0K}^{(T)}$ of the pion exchange terms as
\begin{align}
 f_{\lambda_0K}^{(T)}\!\left(p,p'\right)
 &=
 \frac{3(4\pi)^3}{2}\hat L^2
 \int_{-1}^1 dx P_K(x)
 G^{(T)}(q,Q),
 \label{MPEfunpigeneralT}
\end{align}
with
\begin{align}
 G^{(T)}(q,Q)
 =
 \left\{
 \begin{aligned}
  &-\frac{1}{(2\pi)^3}\!\left(\frac{g_A}{2f_\pi}\right)^{\!2} &&\qquad (\text{$1\pi$}),\\
  &V_{T}^{(2)}(q,Q) &&\qquad (\text{Tensor at NLO}),\\
  &V_{T}^{(3)}(q,Q) &&\qquad (\text{Tensor at N$^2$LO}),\\
  &W_{T}^{(3)}(q,Q) &&\qquad (\text{Tensor with $\tau$ exchange at N$^2$LO}),\\
  &V_T^{(m_N^{-2})}(q,Q) &&\qquad (\text{Tensor for N$^3$LO-$m_N^{-2}$ term}),\\
  &V_T^{(\mathrm{2L})}(q,Q) &&\qquad (\text{Tensor for N$^3$LO-2L term}),\\
  &W_T^{(c_i^2)}(q,Q) &&\qquad (\text{Tensor with $\tau$ exchange for N$^3$LO-$c_i^2$ term}),\\
  &W_T^{(c_i/m_N)}(q,Q) &&\qquad (\text{Tensor with $\tau$ exchange for N$^3$LO-$c_i/m_N$ term}),\\
  &W_T^{(m_N^{-2})}(q,Q) &&\qquad (\text{Tensor with $\tau$ exchange for N$^3$LO-$m_N^{-2}$ term}),\\
  &W_T^{(\mathrm{2L})}(q,Q) &&\qquad (\text{Tensor with $\tau$ exchange for N$^3$LO-2L term}).
 \end{aligned}
 \right.
 \label{GTqQ}
\end{align}
Again, the explicit forms of $G^{(T)}$ can be found in Appendix~\ref{SecChiralpot}.

The $\sigma L$ terms in association with $D_{15}$ and $V_{\sigma L}^{(m_N^{-2})}$ are also the tensor contributions.
The former has the MPE function,
\begin{align}
 f_{L_qL_QK}^{(\sigma L;D_{15})}\!\left(p,p'\right)
 &=
 -864 D_{15}
 \Bigg[\frac{1}{4}\left[\left\{\left(p^2+p'^2\right)^2 -\frac{4}{3}p^2p'^2\right\}\delta_{K0}
 -\frac{8}{3}p^2p'^2\delta_{K2}
 \right]\delta_{L_q 0}\delta_{L_Q 0}
 \nonumber\\
 &+\left[
 \left(p^2+p'^2\right)\delta_{K0}-2pp'\delta_{K1}
 \right]\delta_{L_q 0}\delta_{L_Q 2}
 \nonumber\\
 & +\frac{1}{4}\left[
 \left(p^2+p'^2\right)\delta_{K0}+2pp'\delta_{K1}
 \right]\delta_{L_q 2}\delta_{L_Q 0}
 +\delta_{K0}\delta_{L_q 2}\delta_{L_Q 2}
 \Bigg],
 \label{MPEfunsigmaL}
\end{align}
while that for the latter reads
\begin{align}
 f_{L_qL_QK}^{(\sigma L;m_N^{-2})}\!\left(p,p'\right)
 &=
 -2(12\pi)^3\hat K^2
 \int_{-1}^1 dx P_K(x)
 q^{2-L_q} Q^{2-L_Q}
 V_{\sigma L}^{(m_N^{-2})}(q,Q),
 \label{MPEfunVsigLN3LO}
\end{align}
where $V_{\sigma L}^{(m_N^{-2})}$ is defined by Eq.~\eqref{2piN3LOVsigmaL}.


\bibliographystyle{ptephy}
\bibliography{ChiralAMD}

\end{document}